\documentclass[aps,prb,amsmath,twocolumn,amssymb,superscriptaddress,floatfix,showpacs]{revtex4}

\newcommand{\linecite}[1]{Ref.~\onlinecite{#1}}

\usepackage{color}
\usepackage{ amssymb }
\usepackage{graphicx}
\usepackage{enumerate}
\usepackage{subfigure}
\usepackage{bbold}
\usepackage{verbatim}
\usepackage{float}
\usepackage{dsfont}
\usepackage{multirow}
\usepackage[ocgcolorlinks,colorlinks=true, linkcolor=blue,citecolor=blue]{hyperref}

\begin{document}


\title{Phase diagram of the classical Heisenberg antiferromagnet on a triangular lattice in applied magnetic field}


\author{Luis Seabra}
\affiliation{H.\ H.\ Wills Physics Laboratory, University of Bristol,  Tyndall Av, BS8--1TL, United Kingdom}
\affiliation{Research Institute for Solid State Physics and Optics, H-1525 Budapest, P.O.B. 49, Hungary}
\author{Tsutomu Momoi}
\affiliation{Condensed Matter Theory Laboratory, RIKEN, Wako, Saitama 351-0198, Japan}
\author{Philippe Sindzingre}
\affiliation{Laboratoire de Physique Th\'eorique de la Mati\`ere Condens\'ee, UMR 7600 of CNRS,
Universit\'e P. et M. Curie, case 121, 4 Place Jussieu, 75252 Paris Cedex, France }
\author{Nic Shannon}
\affiliation{H.\ H.\ Wills Physics Laboratory, University of Bristol,  Tyndall Av, BS8--1TL, United Kingdom}


\date{\today}


\begin{abstract}
The Heisenberg antiferromagnet on a two-dimensional triangular lattice is a
paradigmatic problem in frustrated magnetism.
Even in the classical limit $S \to \infty$, its properties are far from simple.
The ``120 degree'' ground state favoured by the frustrated antiferromagnetic
interactions contains a hidden chiral symmetry, and supports two distinct types
of excitation.
And famously, three distinct phases, including a collinear one-third magnetisation
plateau, are stabilised by thermal fluctuations in applied magnetic field.
The questions of symmetry-breaking raised by this model are deep
and subtle, and after more than thirty years of study, many of the details of
its phase diagram remain surprisingly obscure.
In this paper we use modern Monte Carlo simulation techniques to
determine the finite-temperature phase diagram of the classical Heisenberg
antiferromagnet on a triangular lattice in applied magnetic field.
At low to intermediate values of magnetic field, we find evidence for a
continuous phase transition from the paramagnet into the collinear one-third
 magnetisation plateau, belonging to the three-state Potts universality class.
We also find evidence for conventional Berezinskii-Kosterlitz-Thouless
transitions from the one-third magnetisation plateau into the canted
``Y-state'', and into the 2:1 canted phase found at high fields.
However, the phase transition from the paramagnet into the 2:1 canted
phase, while continuous, does not appear to fall into any conventional
universality class.
We argue that this, like the chiral phase transition discussed in zero field,
deserves further study as an interesting example of a finite-temperature
phase transition with compound order-parameter symmetry.
We comment on the relevance of these results for experiments on
magnetic materials with a triangular lattice.
\end{abstract}




\maketitle

\section{Introduction}
\label{section:introduction}

The problem of antiferromagnetism on a triangular lattice occupies a special
place in the history of frustrated magnetism.
The failure of the Ising antiferromagnet on a triangular lattice to order
at {\it any} temperature has been widely celebrated since the pioneering
work of Wannier~\cite{wannier50} and Husimi and Syoji~\cite{husimi50}.
Similarly, the hugely influential idea of a zero-temperature quantum spin-liquid
was first mooted by Anderson in the context of the spin-1/2 Heisenberg antiferromagnet
on a triangular lattice~\cite{anderson73}.
More recent calculations suggest that this model {\it does} order at $T=0$, albeit with a
much-reduced sublattice magnetisation, in a coplanar three-sublattice ``120-degree''
state~\cite{bernu92,capriotti99}.
However, even at a classical level, the finite-temperature physics of this ordered phase is
far from simple.
The 120-degree state possesses a (pseudo)-vector chirality defined by the handedness of the
spin texture in each elementary triangle~\cite{kawamura84b}.
As a consequence, the model can support ${\mathds Z}_2$ vortices as well as conventional
spin-wave excitations, and its low-temperature phase has been argued to be a ``spin gel'' in
which both play an important role~\cite{kawamura84b,kawamura93,kawamura98,okubo10,kawamura10}.
Despite a very determined effort in simulation,
these ideas remain controversial~\cite{southern93,southern95,wintel95,calabrese01,delamotte10}.

Frustrated magnets also exhibit a fantastically rich range of phases in applied magnetic field,
and once again, studies of the triangular lattice antiferromagnet have played a central role in
forming opinion.
The Ising antiferromagnet on a triangular lattice famously exhibits a one-third magnetisation
plateau in applied magnetic field~\cite{mekata77}.
A one-third magnetisation plateau is also found in the Heisenberg antiferromagnet
on a triangular lattice, where it takes the form of a collinear three-sublattice state
stabilised by both thermal~\cite{kawamura85} and quantum~\cite{chubukov89} fluctuations.
Fluctuations select two further phases as a function of magnetic field
--- a coplanar, three-sublattice ``Y--state'', which is a canted version of the 120-degree state,
and a 2:1 canted phase, which is a coplanar, canted version of the one-third magnetisation
plateau~\cite{kawamura85,chubukov89}.
The same succession of phases also occurs in the XY antiferromagnet on a triangular
lattice~\cite{lee86}, and very similar magnetic phase diagrams occur in a wide range of
other models.
Consequently, the magnetisation process of the triangular-lattice antiferromagnet
is often presented as the paradigm for the behaviour of a frustrated magnet under field.
As such it serves as a useful starting point to discuss e.g., the classical
Kagom\'e~\cite{zhitomirsky02,gvozdikova11}, and Shastry-Sutherland~\cite{moliner09}
antiferromagnets.

Further motivation for studying triangular lattice antiferromagnets in field can be
taken directly from experiment.
A full magnetic phase diagram as a function of field and temperature has been
measured for a range of triangular lattice antiferromagnets~\cite{collins97}, which
include the $S$=$1/2$ intermetallic GdPd$_2$Al$_3$~\cite{kitazawa99},
and the $S$=$5/2$ insulating oxides RbFe(MoO$_4$)$_2$~\cite{svistov06}
and Rb$_4$Mn(MoO$_4$)$_3$~\cite{ishii11}.
In all of these cases magnetic anisotropy~\cite{miyashita86} and/or weak interlayer
coupling~\cite{watarai01} must be taken into account.
None the less, clear evidence is found in each case for the three phases found
in a Heisenberg model --- the Y-state, the one-third magnetisation plateau, and the
2:1 canted phase.
More general models, with competing or anisotropic exchange interactions, exhibit even
more complex behaviour as a function of field and temperature.
These remain a subject of intensive study for the examples they provide of novel
magnetic phases~\cite{wang09,alicea09,sen09,heidarian10,seabra10,seabra11,griset11,fishman11,fishman11a},
and for their delicate and subtle finite-temperature phase transitions~\cite{stoudenmire09,melchy09}.

In the context of all this activity, it is perhaps surprising that there are only two published
attempts to determine the phase diagram of classical Heisenberg antiferromagnet on a
triangular lattice in applied magnetic field from Monte Carlo
simulation~\cite{kawamura85,gvozdikova11}.
And, while these authors agree as to the phases present, the
nature of the phase transitions between them remains largely unexplored.
In this paper we attempt to remedy this situation by using modern Monte Carlo simulation
techniques to study the phase transitions which occur in the antiferromagnetic
Heisenberg model on a triangular lattice, as a function of temperature and magnetic field.
The challenge --- and interest --- of this problem stems from the fact that the coplanar
Y-state and 2:1 canted phases break both discrete symmetries of the lattice
{\it and} spin-rotation symmetry in the plane perpendicular to the applied field.
Moreover, since the Mermin-Wagner theorem forbids the breaking of a continuous symmetry
in two dimensions~\cite{mermin66}, spin-rotation symmetry is broken only at the level of a topological,
Berezinskii-Kosterlitz-Thouless (BKT) phase transition~\cite{berezinskii72,kosterlitz73}.
This type of compound symmetry-breaking is notoriously difficult to disentangle in two
dimensions.
The example of coupled Ising [$\mathds{Z}_2$]
and XY [$O(2)$] fields,
in particular, has a long history, dating back to work on spin glasses by Villain~\cite{villain77}.
Questions of $\mathds{Z}_2 \otimes O(2)$ symmetry breaking also arise in
the XY antiferromagnet on a triangular lattice~\cite{miyashita84}, and in
models of coupled Josephson-junction arrays~\cite{teitel83-PRL51}.
A central theme for each of these problems is when --- if ever --- Ising and XY
symmetries are broken in a single, continuous, phase
transition~\cite{choi84,lee91,loison04-book,hasenbusch05,cristofano06,minnhagen07,minnhagen08}.


\begin{figure}[h]
\begin{center}
\includegraphics[width=8cm]{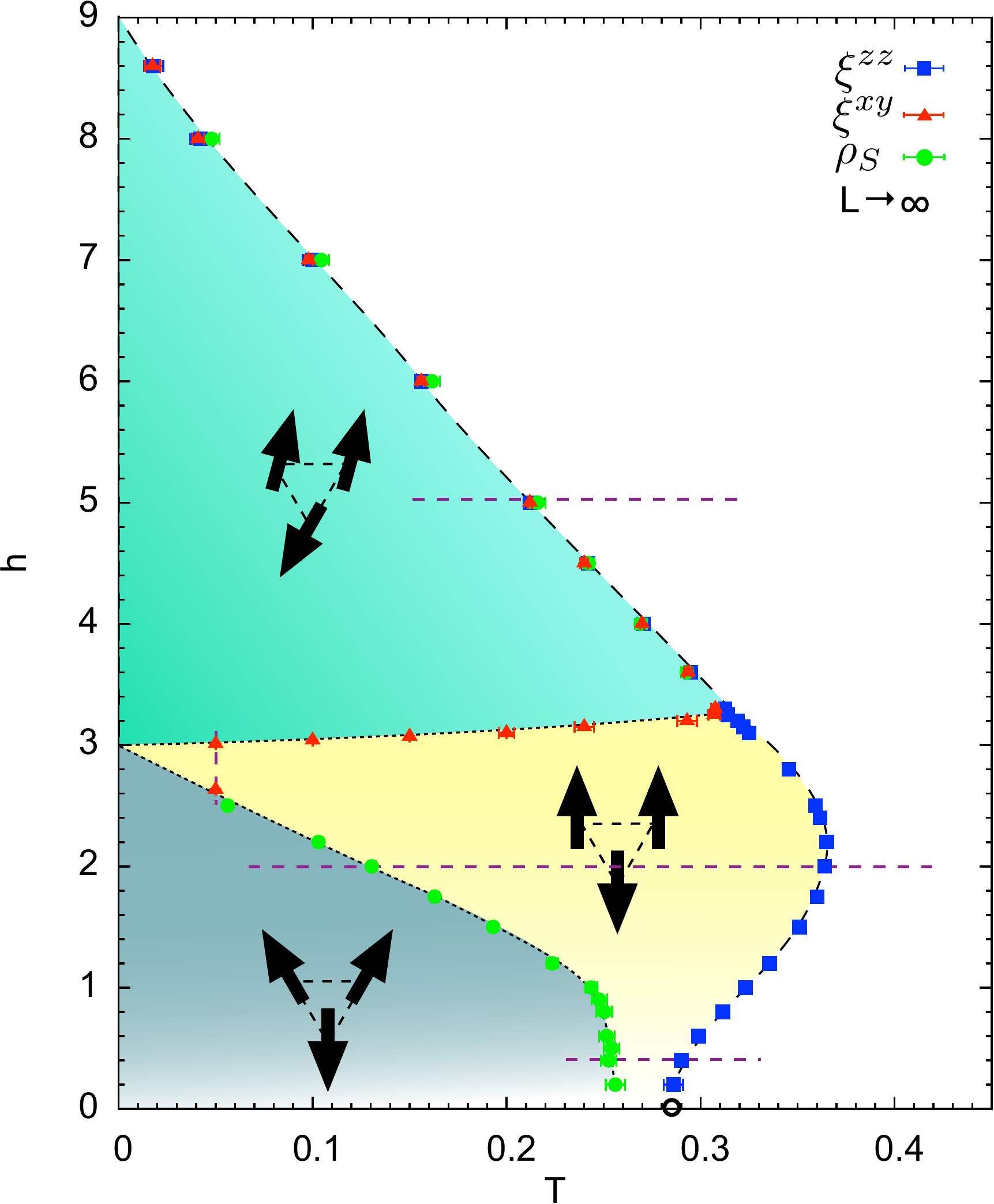}
\end{center}
\caption{
\footnotesize{
(Color online) Magnetic phase diagram of the antiferromagnetic Heisenberg model on a triangular lattice,
obtained from Monte Carlo simulation.
Results have been extrapolated to the thermodynamic limit  using finite-size scaling analysis,
as described in the text.
Continuous phase transitions are drawn with a dashed line, while
Berezinskii-Kosterlitz-Thouless phase transitions are drawn with a dotted line.
For fields $h\lesssim 3$ a double transition is found upon cooling from the paramagnet, while for $h\gtrsim 3$ only
a single transition is found.
Horizontal dashed lines represent cuts at fixed field, $h=0.4$, $h=2.0$ and $h=5.0$, or fixed
temperature $T=0.05$, analysed below.
The low-field region $h \lesssim 0.2$, left unshaded, is beyond the scope of this work.
The black open symbol on the T-axis marks the phase transition found at $h=0$,
from~\linecite{kawamura10}.
}}
\label{fig:phase-diagram}
\end{figure}

In the case of the coplanar phases found in the triangular-lattice Heisenberg antiferromagnet in applied magnetic field,
the relevant symmetry is $\mathds{Z}_3 \otimes O(2)$, and we find that the compound nature
of the order parameters significantly modifies the phase transitions which separate coplanar
phases from the high-temperature paramagnet.
For low values of field, we find a double phase transition, with the system
passing first from the paramagnet into the collinear one-third magnetisation plateau,
and then into the Y-state.
These transitions belong to the three-state Potts and BKT universality classes, respectively.
Another BKT phase transition is found at roughly constant field, separating the one-third
magnetisation plateau and the 2:1 canted state.
Approaching saturation, a \emph{single} continuous phase transition is found from the paramagnet
into the 2:1~canted state.
This exhibits a non-universal jump in spin-stiffness and continuously varying exponents
as a function of magnetic field, and so does not belong to {\it any} conventional universality class.
In reaching these conclusions, we pay careful attention to finite-size effects, which are found
to be very large for low values of magnetic field.
Our results are summarised in Fig.~\ref{fig:phase-diagram}.

The remainder of the paper is structured as follows~:
In Section~\ref{section:methods} we  briefly discuss the Monte Carlo method used and introduce
the order parameters for the different phases, together with associated correlation functions,
and finite-size ansazes.
In Section~\ref{section:overview} we discuss the topology of the overall magnetic phase diagram,
focusing on the importance of finite-size scaling.
In Section~\ref{section:field-sweep} we discuss the transitions
between ordered phases at low temperature.
In Section~\ref{section:low-middle-field} we present representative cases of double phase
transitions upon cooling, characteristic of low and intermediate fields.
In Section~\ref{section:high-field} we discuss a representative case of the single phase transition
observed at high field and its properties.
Lastly, we conclude with an overall summary of our results and a discussion of some of
the remaining open questions in Section~\ref{section:conclusion}.
%

\section{Model, Method and Order Parameters}
\label{section:methods}

The model we consider is defined by the Heisenberg Hamiltonian
\begin{align}
\label{eq:hamiltonian}
\mathcal{H}=J\sum_{\langle i,j \rangle} \mathbf{S}_i \cdot \mathbf{S}_j - h \sum_i S_i^z,
\end{align}
where the sum $\langle i,j \rangle$ runs over all nearest-neighbour bonds
of a triangular lattice (assuming periodic boundary conditions) and the sum on $i$
runs over $N$ lattice sites.
We consider antiferromagnetic exchange interactions $J>0$, and spins are taken
to be classical vectors of unit length.
In the presence of magnetic field it is convenient to rewrite this Hamiltonian as
\begin{eqnarray}
\mathcal{H} &=& \sum_{\triangle} \left[
-\frac{3J}{2} -\frac{h^2}{18 J} + \frac{9J}{2}\left(m - \frac{h}{9J}\right)^2
\right],
\end{eqnarray}
where the sum on $\triangle$ runs over all triangular plaquettes in the lattice and
\begin{eqnarray}
m = \frac{1}{N} \sum_i S^z_i.
\end{eqnarray}
By inspection, for $h < 9J$ the system takes on its minimum energy for
\begin{eqnarray}
m = \frac{h}{9J}.
\label{eq:mofh}
\end{eqnarray}
At zero temperature, this condition selects a manifold of three-sublattice states
which evolves smoothly from the 120-degree state at $h=0$ to saturation at $h=9J$.
In what follows we set $J=1$, and measure $h$ in units of $J$.

The degeneracy of this manifold is lifted by thermal fluctuations, which select collinear and
coplanar states over the non-coplanar ones at low temperature, in a
manifestation of the celebrated ``order-from-disorder'' effect~\cite{kawamura84-JPSJ53}.
The resulting phases are
\begin{enumerate}[i)]
\item a coplanar
Y-state with one spin pinned in the negative $S^z$ direction
and two canting ``up'', i.e. a distorted version of the 120-degree state;
\item a collinear state  at exactly $h=3$,  with two spins ``up'' and one spin ``down'',
i.e. a one-third magnetisation plateau;
\item a coplanar 2:1 canted version of the plateau, which smoothly interpolates
until the collinear saturated paramagnet is reached at $h=9$.
\end{enumerate}
These phases are illustrated in Fig.~\ref{fig:phase-diagram}.

In the collinear one-third magnetisation plateau (ii), only the $S^z$ components of spin
participate in symmetry breaking.
In the presence of magnetic field, this phase breaks only a discrete $C_3  \cong \mathds{Z}_3$
symmetry of the lattice, and long-range order is permitted in 2D.
However, the coplanar Y-state (i) and 2:1 canted phase (iii) also involve spin components in
the $S^x$-$S^y$ plane.
In this case, long-range order implies selecting a common plane for canting, breaking the
$O(2)$ symmetry for rotation of spins about the direction of the magnetic field.
Since the Mermin-Wagner theorem forbids the breaking of this continuous symmetry at
any finite temperature in 2D, only a regime with ``quasi-long-range'' order, described by the
algebraic decay of spin correlations, is permitted.
Both coplanar phases also inherit the broken lattice symmetry, hence  a
compound  $\mathds{Z}_3\otimes O(2)$ symmetry is broken at $T=0$.
These phases with broken mixed symmetries can also be viewed as magnetic
``supersolids''~\cite{matsuda70,liu73,tay10}.


In order to study the finite-temperature properties of the model defined by Eq.~(\ref{eq:hamiltonian})
we perform large-scale parallel tempering~\cite{hukushima96} Monte Carlo simulations.
Simulating this model  is challenging because of the under-constrained nature of the $T=0$ ground state.
At low temperature the system can become frozen in non-coplanar $T=0$ ground states which still obey
the magnetisation constraint, especially for small system sizes.
In this case, the thermal-selection process is not fully realised and strong finite-size effects are visible.
In order to overcome this problem, we couple the parallel-tempering Monte Carlo scheme to successive
deterministic over-relaxation sweeps, which comprise the reflection of each spin around its respective local field.
Since this is a reversible and micro-canonical update, the global Markov chain for parallel tempering and
over-relaxation also obeys detailed balance on the whole.
Simulations  of from 48 to 128 replicas (temperatures) were performed in parallel
for a variety of  \mbox{$L\times L$} rhombohedral clusters with periodic boundary conditions.
The linear size $L$ was chosen to be  commensurate with three-sublattice order in the range $L \in 60-210$.
Typical simulations involved 2$\times$10$^6$ steps, half of which were discarded for thermalisation.
Each step consisted of one local-update sweep of the lattice followed by
two over-relaxation sweeps, with replicas at different temperatures exchanged every 10 steps.
Random initial configurations were employed.


The three ordered phases found break the translational symmetry of the lattice.
This can also be interpreted as the breaking of permutation symmetry between the three different sublattices
in which the triangular lattice can be divided, labelled \emph{A}, \emph{B} and \emph{C}.
In order to study this process, we introduce a complex order parameter
\mbox{$\psi=\psi_{1}+ i \psi_{2}$}, based on a two-dimensional irreducible representation of
the $C_3  \cong \mathds{Z}_3$ lattice rotation group
\begin{align}
\psi_{1}^z &=\frac{3}{\sqrt{6}N} \label{eq:o-c3-1}
\sum_i 2S^z_A +2 S^z_B-4S_C^z, \\
\psi_{2}^z &=-\frac{3}{\sqrt{2}N} \sum_i 2 S^z_B-2S^z_A, \\
|\psi^z|^2 &=|\psi_{1}^z|^2+|\psi_{2}^z|^2, \label{eq:o-c3-3}
\end{align}
where the sum over $i$ runs over the $N/3$ elements of each sublattice $A,B$ and $C$.
Since parallel tempering effectively restores the lattice symmetries, we measure the magnitude
of the order parameter
\begin{align}
\mathcal{O}^{zz}& =\Big<  |\psi^z| \Big>,
\label{eq:c3-aft}
\end{align}
which is normalised to $12/\sqrt{6}$ in the case of a perfect ``two-up, one-down'' collinear configuration
(only achievable at $T=0$ and $h=3$).  The ordering susceptibility and the (temperature-dependent) structure
factor associated with this order parameter are defined~as
\begin{align}
\chi^{zz}& =N\frac{\langle |\psi^z|^2  \rangle - \langle |\psi^z| \rangle^2}{T},
 \label{eq:chi-Z3}\\
\mathcal{S}^{zz}& =N\frac{\langle |\psi^z|^2 \rangle }{T}.
\label{eq:S-Z3}
\end{align}


In order to characterise the phase transitions found we employ the standard finite-size scaling expressions
\begin{align}
\label{eq:m-c3-fss} \mathcal{O}^{zz}&=L^{-\beta/\nu} \tilde{O}^{zz}(tL^{1/\nu}), \\
\label{eq:chi-c3-fss} \chi^{zz}&=L^{\gamma/\nu} \tilde{\chi}^{zz}(tL^{1/\nu}),\\
\mathcal{S}^{zz}&=L^{2-\eta} \tilde{ \mathcal{S}}^{zz} (t L^{1/\nu}),
\label{eq:S-Z3-scaling}
\end{align}
as a function of the reduced temperature $t$=$(T_c-T)/T$.  The critical exponents $\nu$, $\beta$ and $\gamma$
 are obtained from Eqs.~(\ref{eq:m-c3-fss}) and (\ref{eq:chi-c3-fss}), through the usual
data collapse of the respective scaled quantities around the critical point.
Since the scaled structure factor $\tilde{ \mathcal{S}}^{zz}$ becomes independent of system size
exactly at $T_c$,  Eq.~(\ref{eq:S-Z3-scaling}) allows the determination of the correlation-function
exponent $\eta$, provided $T_c$ is found beforehand. 
Error bars for the critical exponents are calculated by assuming a maximum deviation in
 the data points used to perform the data collapse, which results in a conservative (over)
  estimator of the precision in the obtained critical exponents.


The presence of magnetic field reduces the symmetry of the model to \emph{SO(2)} rotations in the
\mbox{$S^x$-$S^y$}~spin plane, which  supports, at most, quasi-long-range order.
The spin stiffness $\rho_S$, which  acts as the (non-local) order parameter for this algebraic order,
can be calculated from the  cost in the free energy of rotating the projection of each spin in the
perpendicular plane $\mathbf{S}_i^\perp=(S_i^x,S_i^y)$, see e.g. \linecite{seabra11} and references therein,
\begin{align}
\rho_s[\textbf{\^{e}}]=&  -\frac{2  }{\sqrt{3}N}  \Bigg\langle J \sum_{{\langle i,j \rangle}}
(\textbf{\^{e}} .\mathbf{r}_{ij})^2  \mathbf{S}^\perp_i.\mathbf{S}^\perp_j
\Bigg\rangle
 \nonumber \\
&
- \frac{2}{\sqrt{3}NT} \Bigg\langle   \Big(  J\sum_{{\langle i,j \rangle}}  (\textbf{\^{e}} .\mathbf{r}_{ij})  \mathbf{S}^\perp_i
 \times \mathbf{S}^\perp_j
 \Big)^2     \Bigg\rangle ,
\label{eq:rhos-3}
\end{align}
where $\mathbf{r}_{ij}$$=$$\textbf{r}_i$$-$$\textbf{r}_j$ and $\rho_S$ has been normalised by $\sqrt{3}/2$ to the unit area.
Since parallel tempering effectively restores the lattice symmetries,  $\rho_S$ is averaged over three
symmetric directions in the lattice
\mbox{$\textbf{\^e}$$=$$(\textbf{\^e}_x,\textbf{\^e}_y)\hspace{-3pt}=\{(1,0),(1/2,\sqrt{3}/2),(-1/2,\sqrt{3}/2)$}\}.


The unbinding of vortex pairs at a BKT phase transition suppresses the spin stiffness,
leading to a jump in $\rho_S$ at the transition temperature $T_{\sf BKT}$.
This jump $\Delta\rho_S$  can be expressed in terms of the correlation-length exponent $\eta$ as~\cite{nelson77}
\begin{eqnarray}
\Delta\rho_S=\frac{T_{\sf BKT}}{2\pi\eta(T_{\sf BKT})}.
\label{eq:rho-jump}
\end{eqnarray}
In a conventional BKT transition, $\eta(T_{\sf BKT}) = 1/4$, and the jump in spin stiffness
takes on the universal value $\Delta\rho_S=2T_{\sf BKT}/\pi$.
However, $T_{\sf BKT}$ is itself strongly renormalised in finite-size simulations, and the correct
ansatz for the finite-size scaling of $T_{\sf BKT}$ must take account of logarithmic
corrections~\cite{weber88}
\begin{align}
T_L=T_{\sf BKT}\Big(1+\frac{1}{2}\frac{1}{\log L + \log b}\Big)
\label{eq:rho-scaling}
\end{align}
Fits of this two-parameter scaling form to finite-size results for the jump in spin stiffness
give an estimate of the true $T_{\sf BKT}$ in the thermodynamic limit.


The perpendicular
component of the $C_3$ order parameter, $\mathcal{O}^{xy}$,
can be defined by analogy with Eqs.(\ref{eq:o-c3-1})-(\ref{eq:c3-aft}).
Since this quantity implies a (staggered) planar magnetisation it must vanish
at any finite temperature in the thermodynamic limit.
However, in a BKT phase transition, the scaling of the perpendicular structure factor $\mathcal{S}^{xy}$,
analogue to Eq.~(\ref{eq:S-Z3}), yields the characteristic critical exponent $\eta(T_{\sf BKT})=1/4$.


Empirically, we find that the analysis of the correlation length $\xi$ provides the most precise method
to obtain the transition temperature for this model, being less sensitive to finite-size effects than,
e.g., Binder cumulants~\cite{binder81}.
In order to calculate this quantity, the structure factor function associated with momentum $\mathbf{q}$
is first defined as
\begin{align}
\mathcal{S}(\mathbf{q})=\Big\langle \frac{1}{N} \Big|\sum_i\mathbf{S}_i\exp(-i\mathbf{q}.\mathbf{r}_i)\Big|^2 \Big\rangle.
\label{eq:corr-fn}
\end{align}
Around the wave vectors corresponding to incipient three-sublattice order ---
\mbox{$\mathbf{q}_{\sf K}=\{ (4\pi/3,0), (2\pi/3,\pi/\sqrt{3}) \}$} --- and in a disordered phase, this quantity displays
the characteristic Lorentzian form $\mathcal{S}(\mathbf{q})\propto\frac{1}{\mathbf{q}^2+\xi^2}$, arising from
short-range correlations.
For a sufficiently large system the correlation length $\xi$ can be obtained from the ratio between
$\mathcal{S}(\mathbf{q})$ at $\mathbf{q}_{\sf K}$ and the nearest allowed wave vector $\mathbf{q}_{\sf K}+ \mathbf{\delta q}$
\begin{align}
\xi=\frac{1}{|\delta \mathbf{q}|}	\sqrt{\frac{\mathcal{S}(\mathbf{q}_{\sf K})}{\mathcal{S}(\mathbf{q}_{\sf K}+\mathbf{\delta q})}-1}.
\label{eq:corr-len}
\end{align}
where $\mathbf{\delta q}=(2\pi/L,0)$.
Eq.~(\ref{eq:corr-len}) is only directly related to the physical correlation length
in the absence of long-range order, that is to say, of Bragg peaks in the structure factor.
The structure-factor function, and thereby the correlation length, can be divided into parallel ($\xi^{zz}$),
and perpendicular ($\xi^{xy}$) components, where the $S^z$ axis is defined by the direction of the magnetic field.


The correlation length becomes infinite at any critical point, whether this is a conventional continuous
phase transition, or the topological transition into a critical BKT phase.
Hence the scaled quantities $\xi^{zz}/L$ or $\xi^{xy}/L$ become independent of system size (or field)
at this temperature, from where  $T_c$ (or $h_c$) can be found in an unbiased way.
Error bars for $T_c$ are estimated using the difference between $T_c$ as obtained by the
 intersection of the scaled $\xi$ between the two largest clusters available and
as obtained the intersection between the second- and third-largest clusters available.
The data collapse of $\xi$ for different system sizes can also be used to extract the correlation-length
exponent $\nu$ at a continuous phase transition
\begin{align}
\xi&=L \tilde{\xi}(tL^{1/\nu}).
\label{eq:xi-scaling}
\end{align}


\section{Topology of the phase diagram}
\label{section:overview}


Our main results are summarised in Fig.~\ref{fig:phase-diagram}.
As mentioned in the introduction, it is not our objective to address the properties of the peculiar
$\mathds{Z}_2\otimes O(3)$ phase transition at $h=0$, although it has been speculated that this
survives the presence of a (very) small external field~\cite{kawamura10}.
As we shall see, accurate simulations become increasingly difficult for very low fields and, owing
to the very large correlation length in the \mbox{$S^x$-$S^y$}~spin plane, require system sizes
larger than the ones presently available.
We will therefore not discuss the phase diagram for $h<0.2$, an area left unshaded
in Fig.~\ref{fig:phase-diagram}.


For $0.2\lesssim h\lesssim 3$, two different phase transitions occur as a function of temperature.
At high temperature a continuous phase transition signals the breaking of  the translational symmetry
of the lattice along the $S^z$ spin direction.
The resulting phase is the collinear one-third magnetization plateau.
We note in passing that the magnetization of this plateau is not tied to one third in a classical
Heisenberg model at finite temperature, since the collinear ``up-up-down''  state is dressed with
thermally excited spin-wave excitations.
(A perfectly collinear  ``up-up-down'' state is realised in a classical model only at $T=0$ and
$h=3$, where it is energetically degenerate with many other, non-coplanar, states).


In the field range $0.2\lesssim h < 3$,  the canted Y-state is found by lowering the temperature
from the one-third magnetization plateau.
The Y-state inherits the broken translational symmetry of the plateau and, since two of its spins are
canted, also breaks spin-rotational symmetry at $T=0$.
At finite temperature, this results in a phase with algebraic order in the $S^x$-$S^y$ plane.
The best interpretation of our numerical results is that, in the thermodynamic limit, the Y-state is
never found to be in contact with the paramagnet, in agreement with other recent works~\cite{gvozdikova11}.


For values of field above the plateau and below the saturation limit, $ 3 \lesssim h < 9 $, a single
\emph{continuous} phase transition separates the paramagnet from the 2:1 canted state.
This transition corresponds to the \emph{simultaneous} onset of long-range order in the $S^z$
spin component and algebraic order in the $S^x$-$S^y$ spin plane.


The transition temperature associated with long-range order along the $S^z$ direction is obtained
by the critical scaling of the corresponding correlation length $\xi^{zz}$.
The transition temperature between the plateau and the Y-state is obtained by the characteristic
BKT finite-size scaling of the jump in the spin stiffness.
Although the critical scaling of the $\xi^{xy}$ correlation length  should yield the same result,
we find that this  method is less  accurate, even for intermediate fields.
This can be explained by the very rapid growth of the $\xi^{xy}/L$ ratio as field is lowered,
until eventually this correlation length exceeds the linear sizes of the available clusters,
rendering this analysis useless.
Hence, as we shall see, the determination of $T_{\sf BKT}$ using $\xi^{xy}$ is less
accurate than that using $\rho_S$ for the same set of system sizes,  even at intermediate fields.


The onset of algebraic order in the $S^x$-$S^y$ plane associated with the 2:1 canted state is
obtained through the onset of critical scaling in $\xi^{xy}$, both as a fixed-temperature scan
from the plateau or as a fixed-field scan from the paramagnetic region.
The transition temperatures obtained with the perpendicular $\xi^{xy}$ and parallel $\xi^{zz}$
correlation lengths agree very well, although they relate to different symmetries.
Moreover, a  jump in the spin stiffness is observed very close to the transition
temperature found with the analysis of the correlation lengths.


\begin{figure}[h]
\begin{center}
\includegraphics[width=8cm]{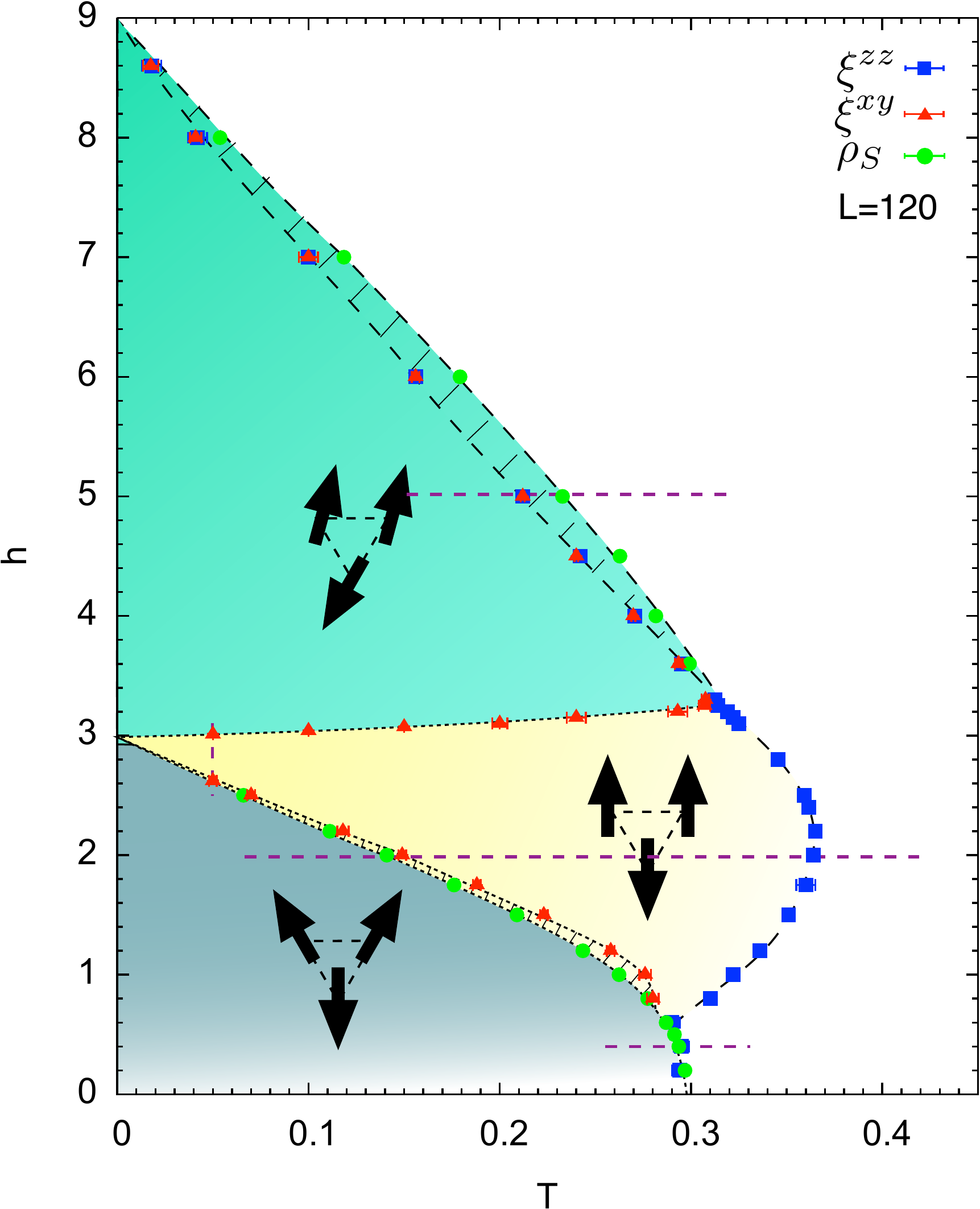}
\caption{\footnotesize{(Color online)
Finite-size pseudo phase-diagram of the antiferromagnetic Heisenberg model on a triangular lattice 
in magnetic field, obtained from Monte Carlo simulation of an $L\times L$ rhombohedral cluster 
with $L=120$.
Continuous ``phase transitions'' are inferred from maxima in the correlation length along the $S^{z}$ 
direction, and drawn with a dashed line.
Berezinskii-Kosterlitz-Thouless ``phase transitions'' in the $S^x$-$S^y$ plane are obtained from 
either the position of the spin-stiffness universal jump for L=120, or from the scaling of the 
corresponding correlation length, and drawn with a dotted line.
A double transition upon cooling is only observed at intermediate fields $0.6\lesssim h \lesssim 3$,~contrary
to the conclusions for $L\rightarrow\infty$,~cf.~Fig.~\ref{fig:phase-diagram}.
A small difference between different measurements of the transitions in the $S^x$-$S^y$ plane is observed for
both low and high fields.
Purple dashed lines indicate cuts at fixed field and temperature analysed below.
The low-field region, $h<0.2$, is beyond the scope of this work.
}}
\label{fig:phase-diagram-L120}
\end{center}
\end{figure}


These conclusions are valid in the thermodynamic limit $L\rightarrow \infty$.
It is instructive to contrast them with the results obtained for a finite-size cluster with $L=120$,
summarised in the pseudo phase-diagram Fig.~\ref{fig:phase-diagram-L120}.
In order to determine the finite-size ``transition'' points we analyse the crossing of the
correlation lengths between a single pair of system sizes, $L=108$ and $L=120$, and
register the temperature where $\rho_S=2T/\pi$ for $L=120$.
At low field, the transitions between the paramagnet and the plateau, and between
the plateau and Y-state, are now indistinguishable, within error bars.
Hence the best interpretation is that a single phase transition separates
the Y-state from the one-third magnetization plateau.
In Fig.~\ref{fig:phase-diagram-L120} the position of the plateau-Y-state transition, obtained
with the critical scaling of  $\xi^{xy}$, is also shown.
The difference between the transition temperature thus obtained and the one obtained using
$\rho_S$ vanishes gradually as field is increased.
Together with the observation that the $\xi^{xy}$ correlation length increases with decreasing
field, eventually becoming larger than  $L=120$ for $h\lesssim0.8$, we can
understand that the determination of $T_{\sf BKT}$ from $\xi^{xy}$ is very strongly affected
by finite-size corrections.
Therefore, we only plot the result obtained with  $\rho_S$ in the $L\rightarrow \infty$  phase
diagram, cf. Fig.~\ref{fig:phase-diagram}.


When passing from the paramagnet to the 2:1 canted state at higher fields,
a similar separation is observed between the transition temperature obtained using the
scaling of correlation lengths, and that obtained from the jump in spin stiffness.
However, once again we observe that this difference vanishes when the correct
finite-size scaling is performed.
We return to this point below.


\section{Field sweep at low temperature}
\label{section:field-sweep}

\begin{figure}[h]
\begin{center}
\includegraphics[width=7cm]{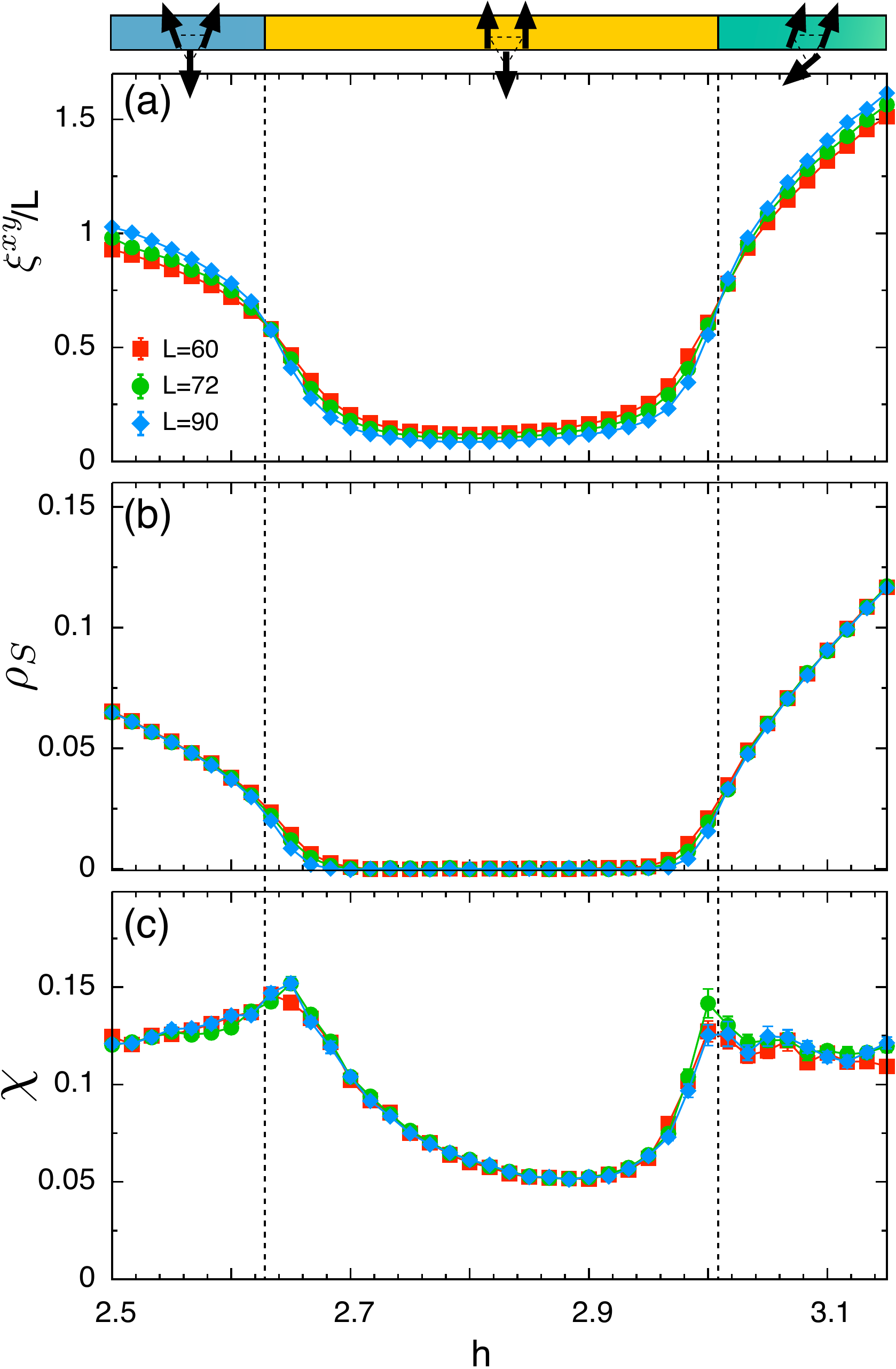}
\end{center}
\caption{\footnotesize{(Color online) Double phase transition as a function of field, at a fixed
temperature $T=0.05$.
The three different ordered phases, Y-state, m=1/3 plateau and 2:1 canted state, are found
with  increasing field.
The magnetisation plateau  is distinguished from the surrounding phases by the suppression
of (a)~the perpendicular correlation length $\xi^{xy}$ and (b) the
 spin stiffness~$\rho_S$.
Its structure is also revealed by the suppression of (c)  magnetic susceptibility $\chi$.
Throughout this paper, lines connecting data points are guides to the eye, unless stated otherwise.
}}
\label{fig:T005}
\end{figure}

We start our analysis of the phase transitions with a scan in field at  fixed temperature
\mbox{$T=0.05$}, cf. Fig.~\ref{fig:T005}, which can also be found as a vertical dashed
line in Fig.~\ref{fig:phase-diagram}.
All the ordered phases discussed are found as the value of field is increased : firstly we
have the low-field Y-state, which is then followed by the one-third plateau at $h=2.633(3)$, and
finally the 2:1 canted state at $h=3.010(3)$.
Leaving the plateau by either lowering or increasing magnetic field corresponds to the
onset of algebraic order in the $S^x$-$S^y$ plane, as observed in the rise of the corresponding
correlation length $\xi^{xy}$ and spin stiffness, cf. Fig.~\ref{fig:T005}(a) and (b).
These simulation runs were performed without parallel tempering,
but the over-relaxation procedure alone was enough to obtain good results.
The transitions in this region are found to be rather easy to simulate, since
correlation lengths are relatively small.


The different nature of the magnetisation plateau, when compared to the surrounding canted
phases,  is clear in the suppression of magnetic susceptibility $\chi$, cf. Fig.~\ref{fig:T005}(c).
This feature becomes more pronounced as the temperature at which the field scan is performed
is lowered.


Both phase boundaries between the plateau and canted phase are approximately linear in
temperature, and can be traced to different spin-wave excitations of the plateau state.
The transition between the plateau and the 2:1 canted state is only weakly dependent on
temperature, indicating that both phases have roughly the same entropy, arising from
similar spin-wave excitations.
The other spin-wave excitation inside the plateau corresponds to a canting of the ``up'' spins.
Since this lowers the total magnetisation along $S^z$, this spin-wave is favoured energetically if field is
decreased, making it more favourable to create a spin-wave than to ``cant'' all spins into the Y-state.
 This ``protects'' the plateau against the decrease of field, and it is this higher entropy of the collinear
 phase~\cite{kawamura84-JPSJ53} which makes the plateau-Y-state transition line slope downwards.


\section{Low and intermediate fields, $h\lesssim 3$}
\label{section:low-middle-field}


\begin{figure}[hb]
\begin{center}
\includegraphics[width=7cm]{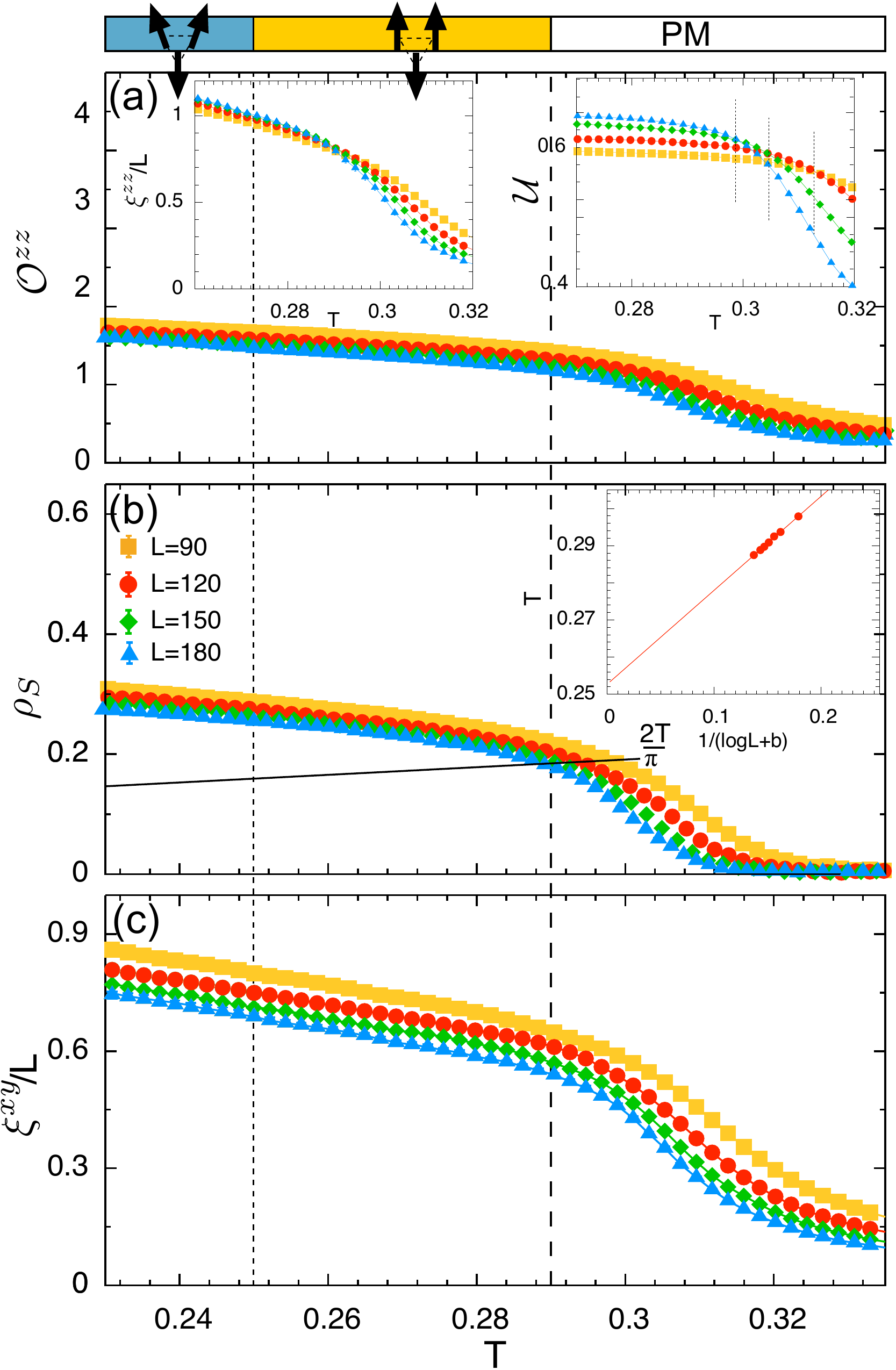}
\end{center}
\caption{\footnotesize{(Color online) Double phase transition as a function of temperature from the 
paramagnet into the plateau and then into the Y-state, at $h=0.4$.
(a) The transition into the plateau breaks a $C_3$ lattice rotation symmetry at a temperature 
$T_{c}=0.290(3)$, identified through the scaling of the correlation length $\xi^{zz}$ [left inset].
The right inset shows the Binder cumulants associated with the $C_3$ order parameter.  
The expected crossing at $T_c$ is subject to very large finite-size effects.
(b) The transition into the Y-state is heralded by the jump in spin stiffness $\rho_S$, 
yielding $T_{\sf BKT}= 0.253(4)$. 
This value is obtained with a $1/\log L$ scaling with system size, as shown in inset to (b).
(c) The perpendicular correlation length $\xi^{xy}$ is so large that no critical crossing is observed for the system sizes studied.
All quantities are strongly renormalised by finite-size effects, even at low temperatures.
 }}
\label{fig:h=04}
\end{figure}


For values of applied field in the range $0.2\lesssim h < 3$ a double phase transition is found
as temperature is lowered from the paramagnetic region.
In Fig.~\ref{fig:h=04} we present results for $h=0.4$, which is representative of the low-field
region $0.2\lesssim h \lesssim1.2$.
The low-temperature region displays a finite value of the $C_3$ order parameter
$\mathcal{O}^{zz}$ [Eq.~(\ref{eq:c3-aft})], cf.~Fig.~\ref{fig:h=04}(a), and spin stiffness
$\rho_S$ [Eq.~(\ref{eq:rhos-3})], cf.  Fig.~\ref{fig:h=04}(b).
This is indicative of, respectively, long-range order in the direction parallel to field and
quasi-long-range (algebraic) order in the perpendicular plane, as expected in the Y-state.
However, the absolute value of these quantities at finite temperature is \emph{strongly}
dependent of the system size. This can be attributed to the proximity of the $h=0$ point,
where both these order parameters vanish and the correlation length is very large
(but probably finite)~\cite{kawamura10}.
This explains the unusually strong finite-size corrections, which vanish gradually as
field is increased.
In fact, the most spectacular demonstration of these problems lies in the  absence of
critical scaling of the $\xi^{xy}$ correlation length, i.e. there is no common crossing or
collapse for different  system sizes, cf. Fig.~\ref{fig:h=04}(c).
Since  there is no long-range order in the $S^x$-$S^y$ plane, Eq.~(\ref{eq:corr-len})
still provides an accurate estimation of the $\xi^{xy}$ correlation length in that region
(obviously, the same no longer holds for~$\xi^{zz}$).
The absence of a merger, or even a crossing, implies that the asymptotic regime, where 
the correlation length is infinite (i.e. $\xi^{xy}/L\approx1$ for finite clusters), has not been reached in the lattice 
sizes studied.   This may be due to a slowdown in simulation dynamics, arising from the
 pathological properties of the $h=0$ point~\cite{kawamura10}.
For higher values of magnetic field $h \approx 0.8$, a critical crossing is only observed
in the largest pair of system sizes studied, $L=180$ and $L=210$.
With increasing magnetic field this crossing  is observed for gradually decreasing system sizes.


These  unusually strong finite-size effects make the accurate determination of the transition
temperatures very hard.
For $h=0.4$, the critical scaling of the  $\xi^{zz}$ correlation length  yields a reasonably
well-converged value (i.e. the movement of the crossing point between successive system
sizes becomes smaller and is not significant for the largest ones employed)
at $T_c=0.290(3)$, cf. left inset to Fig.~\ref{fig:h=04}(a).
This phase transition corresponds to the breaking of translational symmetry.
We reserve its characterisation to  later in the text, for a value of field that allows a cleaner
interpretation.
The determination of $T_c$ through the analysis of the Binder cumulants for the two-component $\mathcal{O}^{zz}$ order parameter, \mbox{$\mathcal{U}=1-\frac{\langle |\psi^z|^4\rangle}{2\langle |\psi^z|^2\rangle^2}$}, converges noticeably slower to the thermodynamic limit,  cf. right inset to Fig.~\ref{fig:h=04}(a).

\begin{figure}[b!]
\begin{center}
\includegraphics[width=7cm]{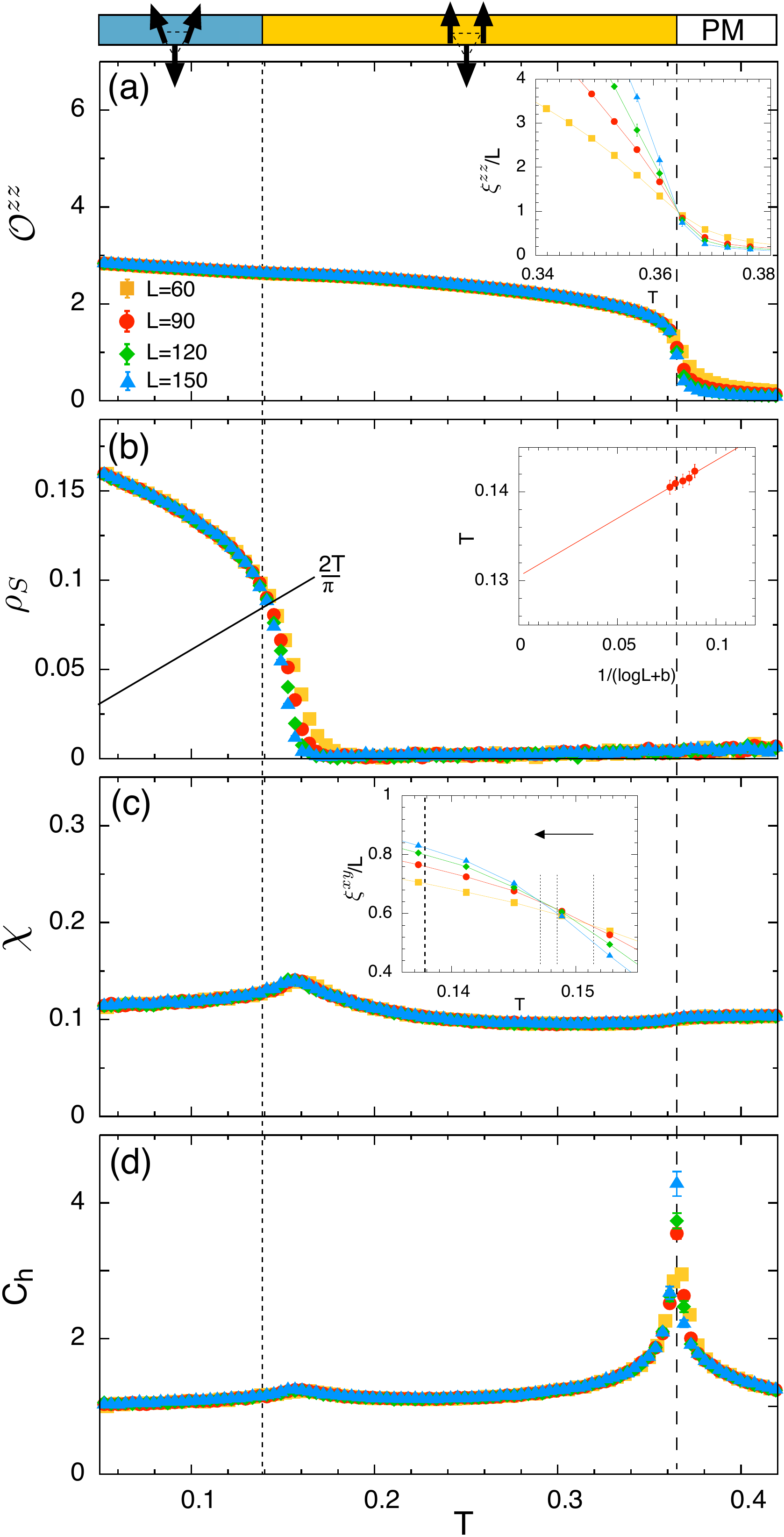}
\end{center}
\caption{\footnotesize{(Color online)
Double phase transition as a function of temperature from the paramagnet state into
the collinear $m$=1/3 plateau and then into the coplanar Y-state, at $h=2$.
(a) The transition into the plateau is heralded by the rise in the  $C_3$  order parameter.
The transition temperature is $T_c=0.364(1)$, obtained with the critical scaling of the correlation
length $\xi^{zz}$ in inset to (a).
(b) The spin-stiffness $\rho_S$ jump signals the BKT transition into the Y-state at
$T_{\sf BKT}= 0.138(3)$, a value obtained with a logarithmic scaling with system size,
as shown in inset to (b).
The paramagnet-plateau transition only displays a weak signature in (c) magnetic
susceptibility, but shows a clear peak in (d) heat capacity.
The position of the inner magnetic susceptibility peak in (c), scaling of  the correlation
length $\xi^{xy}$ in inset to (c), and the inner heat capacity peak in (d) give a inaccurate
estimative of $T_{\sf BKT}$.  }}
\label{fig:h=2-I}
\end{figure}



Analysing now the $S^x$-$S^y$ spin-texture plane, the best fit to the evolution with system
size of the position of the universal jump in spin stiffness is given by a logarithmic form
[Eq.~(\ref{eq:rho-scaling})], with
$T_{\sf BKT}=0.253(4)$ and $b=-1.7084(5)$,  cf. inset to Fig.~\ref{fig:h=04}(b).
The resulting value for $T_{\sf BKT}$ is \emph{significantly} lower than $T_c$, clearly implying
an intermediate phase between the paramagnet and the \mbox{Y-state} that only breaks
translational symmetry, i.e. the one-third magnetization plateau.
A  jump in spin stiffness at $T\approx0.25$ in the thermodynamic limit implies
a remarkable finite-size renormalisation of spin-stiffness, since $\rho_S$ is finite
for the lattice sizes studied spin stiffness in a broad region above that temperature,
cf. Fig.~\ref{fig:h=04}(b).
This agrees with the strong variation of $\rho_S$ value with system size, as observed
even deep inside the Y-state.
It should be emphasised that the separation of these two transitions is observed
only in the $L\rightarrow \infty$ limit.
Taken at a fixed system size, both estimates for the transition temperature coincide 
at $T_c\approx 0.291(4)$ (within error bars) for the finite-size systems studied, 
cf. Fig.~\ref{fig:phase-diagram-L120}.
It is tempting, therefore, to infer that these two transitions appear to take
place at the same temperature.
However, since the correlation length is very large, any comparison
 at a fixed finite size is unreliable.


A further manifestation of strong finite-size effects is seen in the presence of a small
amount of non-coplanarity in the configurations found in equilibrium.
This can be observed by looking at e.g. scalar chirality or quadrupolar spin moment quantities, which
show a distinct signature inside the $Y$-phase [not shown].
These signals of non-coplanarity scale to zero with increasing system size, but very slowly.
In spite of all  these problems, the value of $T_{\sf BKT}$ extracted from the
jump in spin stiffness $\rho_S$, obeys the logarithmic evolution with system size expected
for a BKT transition for fields as low as $h=0.2$ [see inset to Fig.~\ref{fig:h=04}(c)].



In order to obtain a clean characterisation of this double phase transition, we perform a
similar analysis for a value of field $h=2$, where these two transitions are now
well separated, cf. Fig.~\ref{fig:h=2-I}.
A sharp rise of the $C_3$ order parameter, cf. Fig.~\ref{fig:h=2-I}(a), is associated with the onset
of long-range order when entering the magnetisation plateau.
This is observed to happen at $T=0.364(1)$,
as found in the critical scaling of $\xi^{zz}$, cf. the inset to Fig.~\ref{fig:h=2-I}(a).
This phase transition can also be observed in thermodynamic signatures such as
a very shallow suppression of the magnetic susceptibility [Fig.~\ref{fig:h=2-I}(c)], and a
sharp peak in heat capacity [Fig.~\ref{fig:h=2-I}(d)].

\begin{figure}[h]
\begin{center}
\includegraphics[width=9cm]{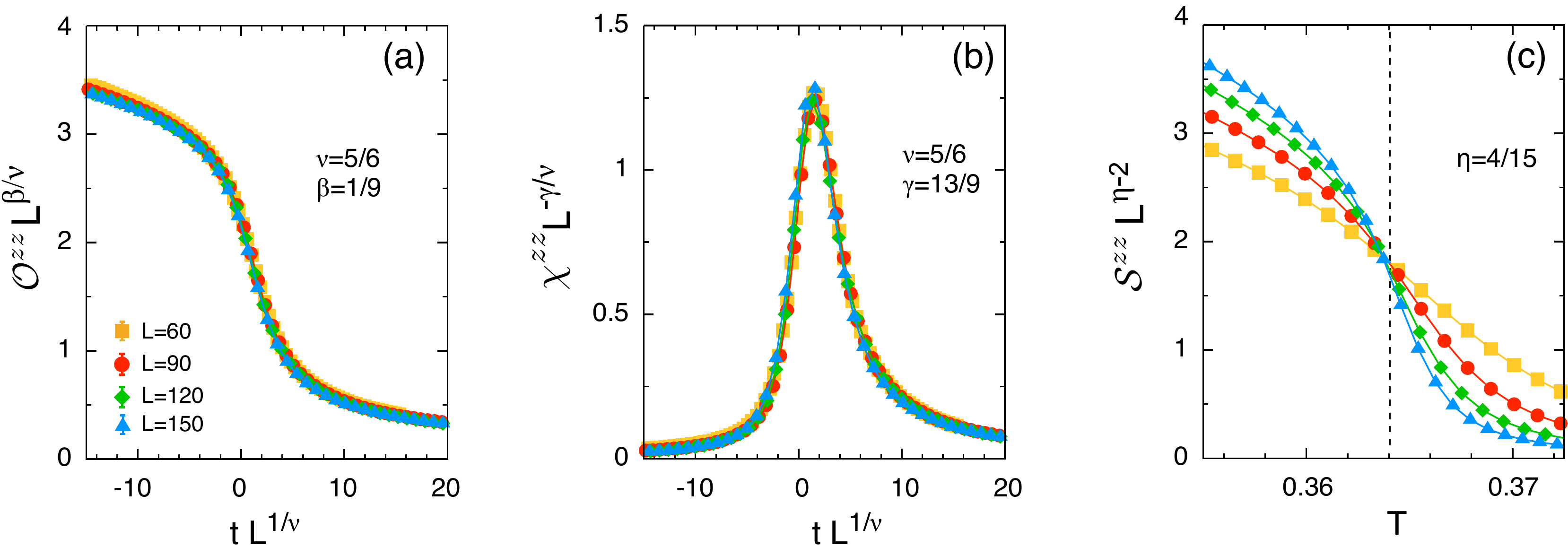}
\end{center}
\caption{\footnotesize{(Color online) Data collapse for the continuous phase transition
at $h=2.0$, $T=0.364$, using the exact three-state Potts universality class critical exponents.
 (a) $C_3$ order parameter, (b) $C_3$ order parameter susceptibility and
 (c) associated structure factor $\mathcal{S}^{zz}$. }}
\label{fig:h=2-II}
\end{figure}

A finite-size scaling analysis of this phase transition is performed using the critical exponents
for the three-state Potts model in two dimensions 
\mbox{$\nu=5/6$} (correlation length),
\mbox{$\beta=1/9$} (order parameter),
\mbox{$\gamma=13/9$}  (order parameter susceptibility)
and \mbox{$\eta=4/15$} (correlation function) [\linecite{wu82}].
A perfect data collapse is obtained for the order parameter [Fig.~\ref{fig:h=2-II}(a)],
order-parameter susceptibility [Fig.~\ref{fig:h=2-II}(b)],
and structure factor at \mbox{$T=T_c$} [Fig.~\ref{fig:h=2-II}(c)].
This unambiguously confirms that this phase transition belongs to the three-state
Potts universality class for $h=2.0$.


At a lower  temperature $T\approx 0.15$, the weak features in heat capacity and magnetic
susceptibility herald a transition associated with quasi-long-range ordering in the $S^x$-$S^y$
plane.
This can be interpreted  as the formation of  vortex pairs in the spin texture defined by the
$S^x$-$S^y$ plane, arising from the two canted spins of the Y-state configuration.
Therefore we find a BKT transition and respective rise in the spin stiffness, cf.  Fig.~\ref{fig:h=2-I}(b).
The transition temperature is found by tracking the position of the universal jump in the spin stiffness,
$T_{\sf BKT}$ in  inset to Fig.~\ref{fig:h=2-I}(b).
 Once again the best fit to the finite-size scaling is given by a logarithmic function  of system size,
 yielding $T_{\sf BKT}=0.138(3)$ and $b=1.5051(4)$.

\begin{figure}[h]
\begin{center}
\includegraphics[width=2.9cm]{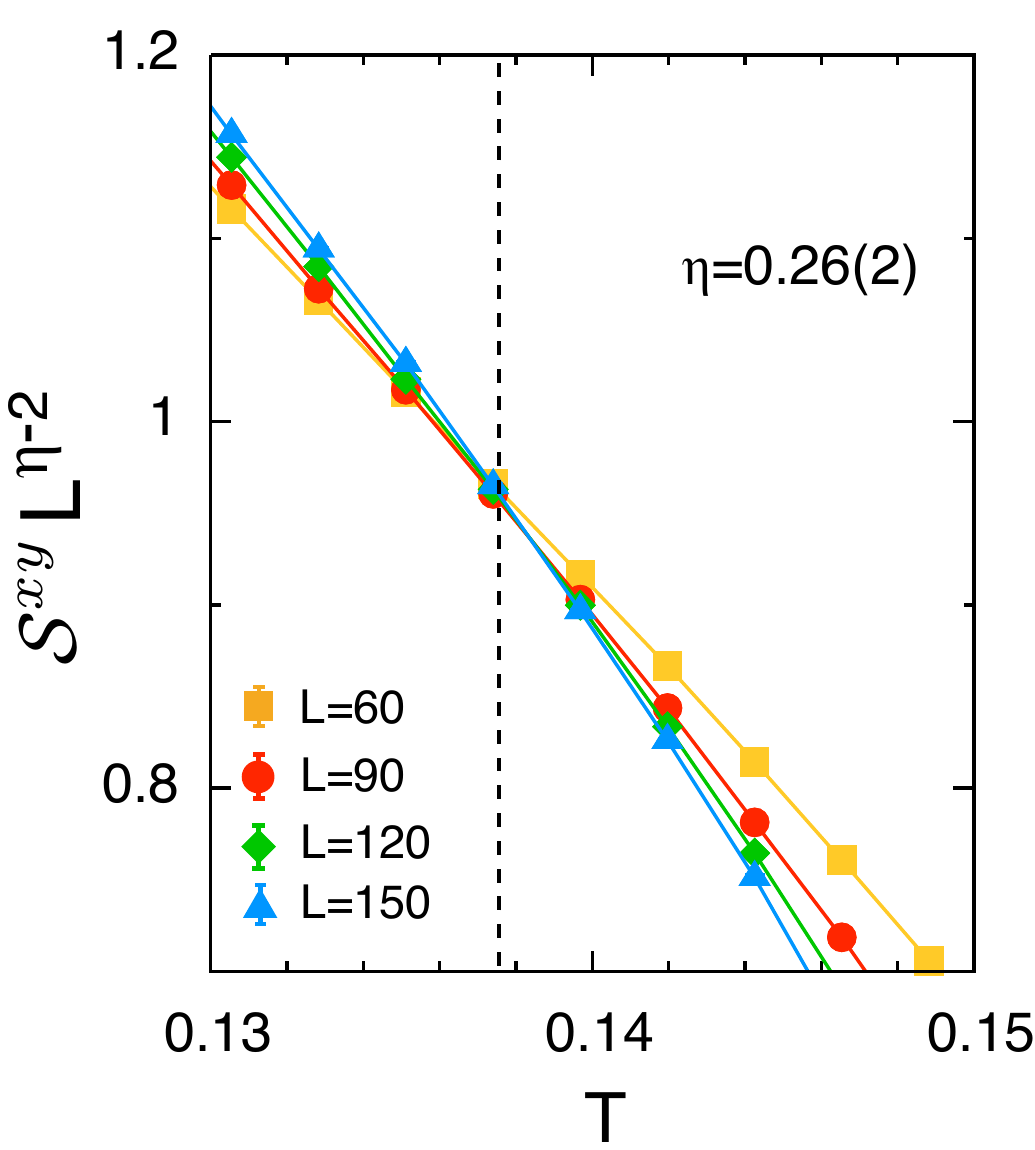}
\end{center}
\caption{\footnotesize{(Color online)
Evidence for Berezinskii-Kosterlitz-Thouless nature of the lower transition at \mbox{$h=2.0, T=0.138$}.
The collapse of the structure factor $\mathcal{S}^{xy}$ at the critical critical temperature from
Fig.~\ref{fig:h=2-I}(b) yields a critical exponent \mbox{$\eta=0.26(2)$}, in good agreement with the
BKT universality class.
}}
\label{fig:h=2-III}
\end{figure}

In order to confirm the BKT character of this transition, we study  the critical scaling of
$\mathcal{S}^{xy}$, the structure factor associated with the  component of the $C_3$
order parameter perpendicular to field.
The crossing of $\mathcal{S}^{xy}/L^{2-\eta}$ for different system sizes
at  $T_{\sf BKT}$ (using the value found with the $\rho_S$ analysis) occurs for a value of the
correlation exponent $\eta=0.26(2)$, cf.~Fig.~\ref{fig:h=2-III}.
This is in good agreement with $\eta(T_{\sf BKT})=1/4$, the expected value in a
standard BKT transition.


If we accept this finite-size scaling analysis at face value, we are lead to the conclusion that,
for all values of magnetic field $0.2 \lesssim h < 3$, the $\mathds{Z}_3$ and $O(2)$ symmetries
are broken at different temperatures.
These two, distinct phase transitions are themselves perfectly conventional.
The phase transition from the paramagnet to the collinear one-third magnetization plateau
shows three-state Potts character, while the phase transition from the one-third magnetization
plateau to the algebraically correlated Y-state is BKT in nature.
This is the interpretation given in the phase diagram Fig.~\ref{fig:phase-diagram}.


However this interpretation needs to be approached with some caution.
Even for $h=2$, some of the quantities calculated are strongly affected by finite-size
corrections.
This can be readily observed in the inset of Fig.~\ref{fig:h=2-I}(c), where it is shown that the
value of $T_{\sf BKT}$ obtained from the critical scaling of the transverse correlation
length $\xi^{xy}$ depends on the size of the lattice studied.
The absence of a good data collapse within the algebraically correlated
Y-state is further evidence of strong finite-size effects.
It would therefore be premature to rule out a single phase transition for $h \to 0$.
Given that the clusters used in the present simulations are not small
(180$\times$180 = 32,400 sites), a fairly heroic act of simulation may be
needed to finally resolve this question.


\section{High Field, $h\gtrsim 3$}
\label{section:high-field}



\begin{figure}[hp!]
\begin{center}
\includegraphics[width=7.2cm]{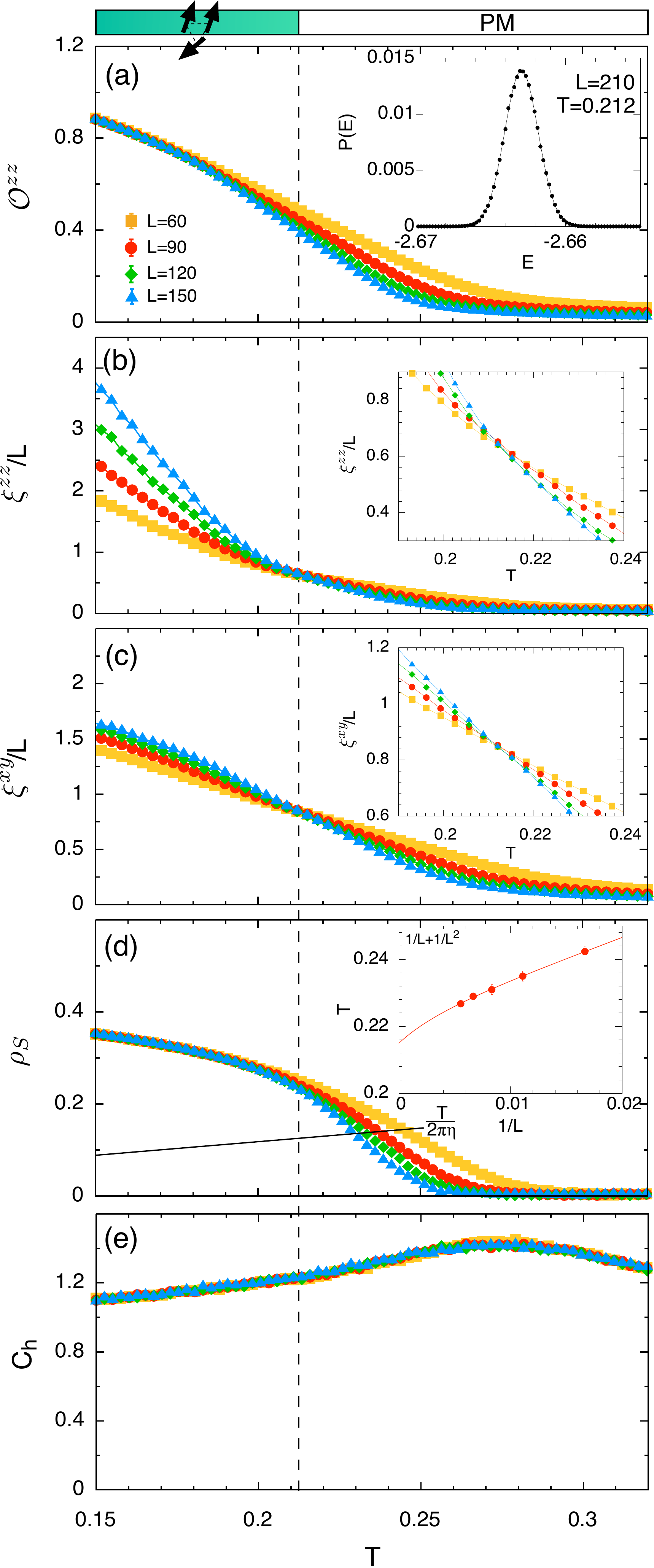}
\end{center}
\caption{\scriptsize{(Color online)
Continuous phase transition from the paramagnet into the 2:1 canted state
as a function of temperature, for $h=5$.
(a)  $C_3$  order parameter, measuring the broken translational symmetry.
Inset to (a) shows
the single-peaked energy distribution  at the transition temperature.
(b) Scaling of the parallel correlation length  $\xi^{zz}$, showing strong  finite-size effects above the transition.
 Inset to (b) shows a detail of the scaling of $\xi^{zz}$ which yields a crossing at $T_c=0.212(1)$.
(c) The scaling of the perpendicular   correlation length $\xi^{xy}$ also shows strong finite size effects above $T_c$. This
transition temperature is also found to be $T_c=0.212(1)$, as shown in the inset to (c).
 (d) The rise in spin stiffness $\rho_S$ also heralds the entry into the 2:1 canted state.
In the inset to (d) a  $1/(2\pi\eta(T_c))$ jump of $\rho_S/T$ is scaled as a function of $1/L+1/L^2$
resulting in $T_{c}= 0.217(4)$, in relatively good agreement with the scaling of the
correlation lengths presented in insets to (b) and (c).
(e) Heat capacity shows a broad peak at an higher temperature.
}}
\label{fig:h5-I}
\end{figure}


The behaviour of the Heisenberg antiferromagnet on a triangular lattice for magnetic field
$h>3$ is dramatically different.
The zero-temperature state is a 2:1 canted version of the one-third magnetization plateau,
which breaks the translational symmetry of the lattice and  spin-rotational symmetry in the
$S^x-S^y$ plane (at $T=0$), i.e the same symmetries as the Y-state studied above.
However,  a  \emph{single} phase transition mediates between the high-temperature paramagnet
and the 2:1 canted phase, in clear contrast with the case described in Section~\ref{section:low-middle-field}.

A selection of results for a representative field value $h=5$ is shown in Fig.~\ref{fig:h5-I}.
The rise of the $C_3$~order parameter, in Fig.~\ref{fig:h5-I}(a), heralds  the onset of long-range
order along the $S^z$ direction.
Strong finite-size artifacts are observed in the paramagnetic region close to the transition,
but the scaling of the $\xi^{zz}$ correlation length  yields a well-converged value of
\mbox{$T_c=0.212(1)$}, cf. Fig.~\ref{fig:h5-I}(b),  with negligible finite-size corrections.
The critical scaling of the perpendicular correlation length  $\xi^{xy}$, cf. Fig.~\ref{fig:h5-I}(c),
results in a value of $T_c=0.212(1)$ (again with negligible finite-size corrections) which is in
perfect agreement with the value obtained with $\xi^{zz}$.
We observe that this very good agreement, smaller than the statistical error bars
$\Delta T=0.001-0.004$, is achieved for all values of fields $h\gtrsim 3.3$.
More precisely, this agreement implies that the correlations along the $S^z$ direction
and the $S^x$-$S^y$ plane become critical at the same temperature.
This is good evidence for a single phase transition into the Y-state, without any
intermediate phase.
In the absence of any symmetry-breaking field in the $S^x$-$S^y$ plane, the rise of the
respective correlation length corresponds to the onset to algebraic order, cf. Fig.~\ref{fig:h5-I}(c).
This can also be observed in the rise of spin stiffness, cf.~Fig.~\ref{fig:h5-I}(d).


Such a phase transition, which breaks a compound symmetry $\mathds{Z}_3\otimes O(2)$,
need not show the behaviour expected of either a three-state Potts, or BKT transition.
This phase transition retains a continuous character up to the largest cluster  size studied $L=210$,
as can be verified by the unimodal energy distribution at the calculated $T_c$, cf. inset to Fig.~\ref{fig:h5-I}(a).
We have also explicitly checked that the Binder cumulant for energy does not develop any
characteristic signatures of a first-order transition as $L\rightarrow \infty$ [not shown].
The heat capacity, cf. Fig.\ref{fig:h5-I}(e), only shows a broad peak at an higher temperature than
the estimated $T_c$.
Although the peak does not diverge with increasing  system system size, the temperature of its
maximum becomes lower.

\begin{figure}[h!]
\begin{center}
\includegraphics[width=9cm]{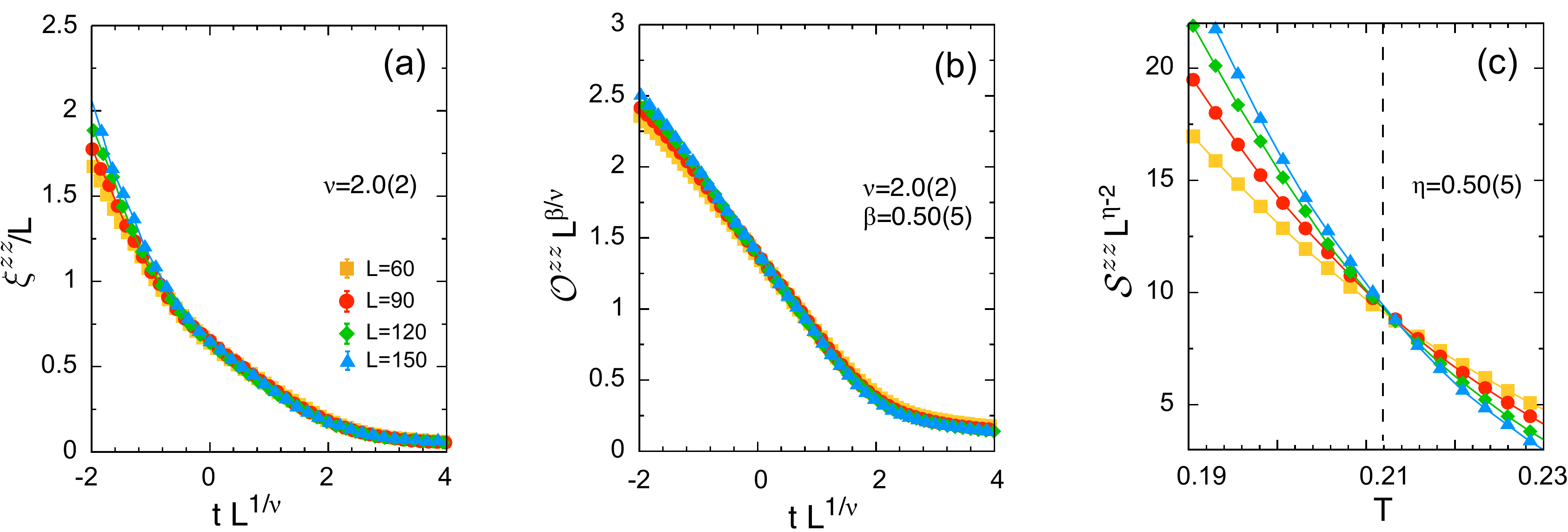}
\end{center}
\caption{\footnotesize{(Color online)
Scaling analysis of the continuous phase transition between the paramagnet and 2:1
canted state at $h=5,T=0.212$.
(a) The collapse of simulation results for the $\xi^{zz}$ correlation length yields the correlation length
exponent $\nu=2.0(2)$.
(b) The scaling of the respective component of the order parameter parallel gives $\nu=2.0(2)$ and $\beta=0.50(5)$.
(c) The critical scaling of the parallel $\mathcal{S}^{zz}$  structure factor gives the correlation exponent $\eta=0.50(5)$.
 }}
\label{fig:h5-II}
\end{figure}

The strong finite-size corrections, observed at $T>T_c$ in e.g. Fig.~\ref{fig:h5-I}(a),
make the finite-size scaling analysis much less precise than in Section~\ref{section:low-middle-field}, but
some conclusions can still be reached.
For $h=5$, it is possible to reliably extract the correlation-length
exponent $\nu=2.0(2)$, the order-parameter exponent $\beta=0.50(5)$, and the correlation-function
exponent $\eta=0.50(5)$  along the $S^z$ spin direction, cf. Fig.~\ref{fig:h5-II}.
This combination of critical exponents does not appear to belong to any known universality class.


\begin{figure}[h!]
\begin{center}
\includegraphics[width=6cm]{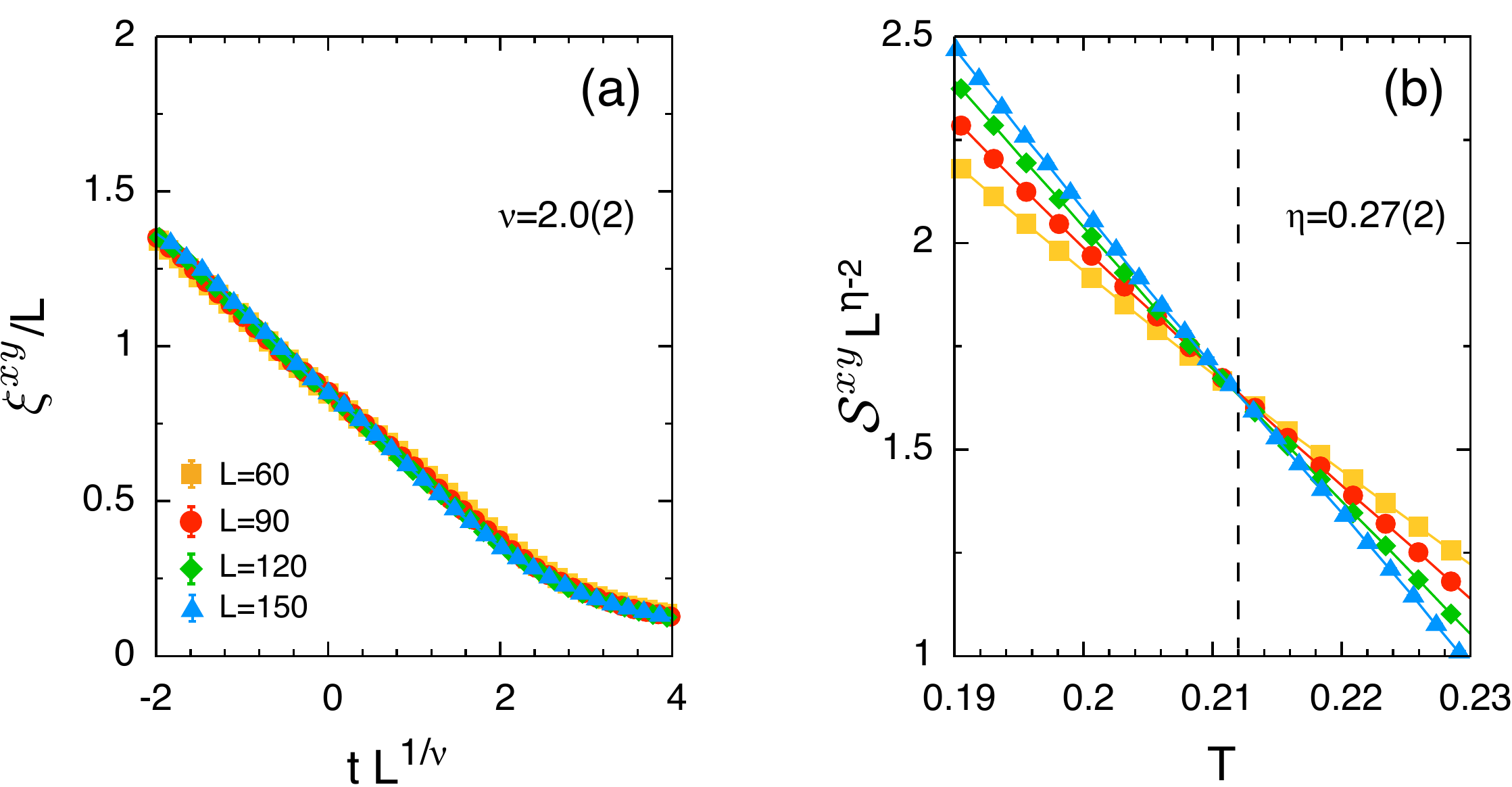}
\end{center}
\caption{\footnotesize{(Color online)
Scaling analysis of the continuous phase transition between the paramagnet and 2:1
canted state at $h=5,T=0.212$.
(a) The collapse of the $\xi^{xy}$ correlation length yields the correlation length
exponent $\nu=2.0(2)$, exactly as in the $S^{zz}$ direction, cf. Fig.\ref{fig:h5-II}(a).
(b) The collapse of the $\mathcal{S}^{xy}$ structure factor gives the
correlation exponent $\eta=0.27(2)$.
 }}
\label{fig:h5-III}
\end{figure}

The finite-size scaling analysis of the perpendicular component of the correlation length $\xi^{xy}$
results in $\nu=2.0(3)$, cf.~Fig.~\ref{fig:h5-III}(a). This closely matches the $\nu$ exponent obtained
for the parallel component of the correlation length.
This is in contrast with the BKT universality class, where the correlation length diverges exponentially.
The value obtained for the correlation function
exponent $\eta=0.27(2)$ at $T=0.212$, our estimate for $T_c$, cf.~Fig.~\ref{fig:h5-III}(b), is
interestingly close to what is expected in a BKT transition.
Nevertheless, this is, apparently, just a coincidence, as we shall see.
%

We anticipate that, regardless of the details of the phase transition, the low temperature
2:1 canted phase will contain bound pairs of vortices in the spin texture, which will
unbind at $T=T_c$, and precipitate a jump in the spin stiffness $\rho_S$.
However the fact that generically $\eta \ne 1/4$ means that the jump in the spin stiffness
need not have the ``universal'' value $\Delta\rho_S = 2T/\pi$ [cf. Eq.~(\ref{eq:rho-jump})].


This interpretation is corroborated by the analysis of the data in Fig.~\ref{fig:h5-I}(d) and
its inset --- the extrapolation to $L\rightarrow \infty$ of the temperature at which the
$\frac{T}{2\pi\eta}$ jump occurs yields the value $T_c=0.217(4)$.
This value agrees within errors with the estimate $T_c=0.212(1)$
obtained with the scaling of the correlation lengths.
The best fit to the finite-size data is now given by the power-law form $a/L+b/L^2$, \emph{not}
the $1/\log(L)$ scaling expected in a BKT transition.
However if this transition does not belong to the BKT universality class, there is no good
\emph{a priori} reason to assume that $T_c$ scales logarithmically with system size.


The $h=5$ results presented above are broadly representative of the phenomenology of the
single transition from the paramagnet into the 2:1 canted state, in the range $3\lesssim h <9 $.
However, important details such as the critical exponents change as a function of field.



\begin{figure}[h]
\begin{center}
\includegraphics[width=7cm]{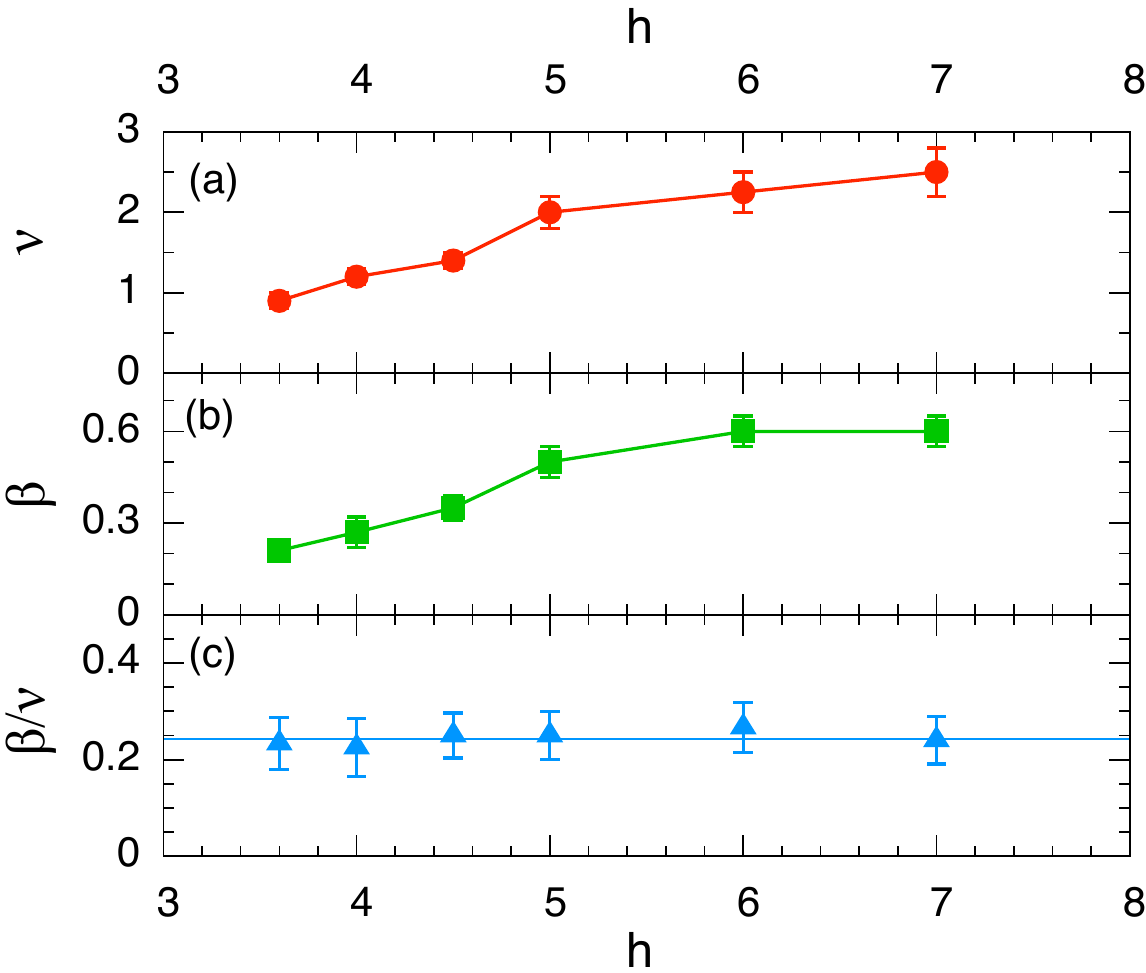}
\end{center}
\caption{\footnotesize{(Color online)
Evolution of the critical exponents associated with correlations of the $S^z$ components
of spin at the continuous phase transition between the paramagnet and 2:1 canted state,
as a function of magnetic field $h$.
Both (a) the correlation-length exponent $\nu$  and (b) the order-parameter
exponent $\beta$ increase with increasing magnetic field --- lines are guides to the eye.
However, (c) the ratio $\beta/\nu$  is roughly constant at $0.24(3)$ --- horizontal line.
 }}
\label{fig:exponents}
\end{figure}


Both $\nu$ and $\beta$ exponents, associated with the $S^z$ component, increase monotonically
with magnetic field, as can be observed in Fig.~\ref{fig:exponents}(a) and (b).
However, the $\beta/\nu$ ratio remains roughly constant at \mbox{$\beta/\nu=0.24(3)$},
cf. Fig.~\ref{fig:exponents}(c).
The two-dimensional scaling law \mbox{$\eta=2\beta/\nu$} therefore implies \mbox{$\eta\sim0.5$},
which agrees  with  the result from the scaling of the structure factor at $T_c$ for the studied fields, as exemplified
in Fig.~\ref{fig:h5-II}(c) for $h=5$.
The order-parameter susceptibility $\chi^{zz}$ is strongly affected by finite-size effects (not shown). 
However, the critical scaling of  $\chi^{zz}$  yields a value of roughly $\gamma\approx 3$ for $h=5$, which is consistent with the scaling relation \mbox{$\gamma=\nu(2-\eta)$}.
The hyper-scaling law in two dimensions $\alpha=2(1-\nu)$, combined with Fig.~\ref{fig:h5-II}(a),
means that the heat-capacity exponent $\alpha$ is negative for this phase transition.
This explains the absence of a singularity in specific heat at the transition temperature
[cf Fig.~\ref{fig:h5-I}(e)], and clearly distinguishes the continuous phase transition here found
from a three-state Potts transition.

\begin{figure}[h]
\begin{center}
\includegraphics[width=7cm]{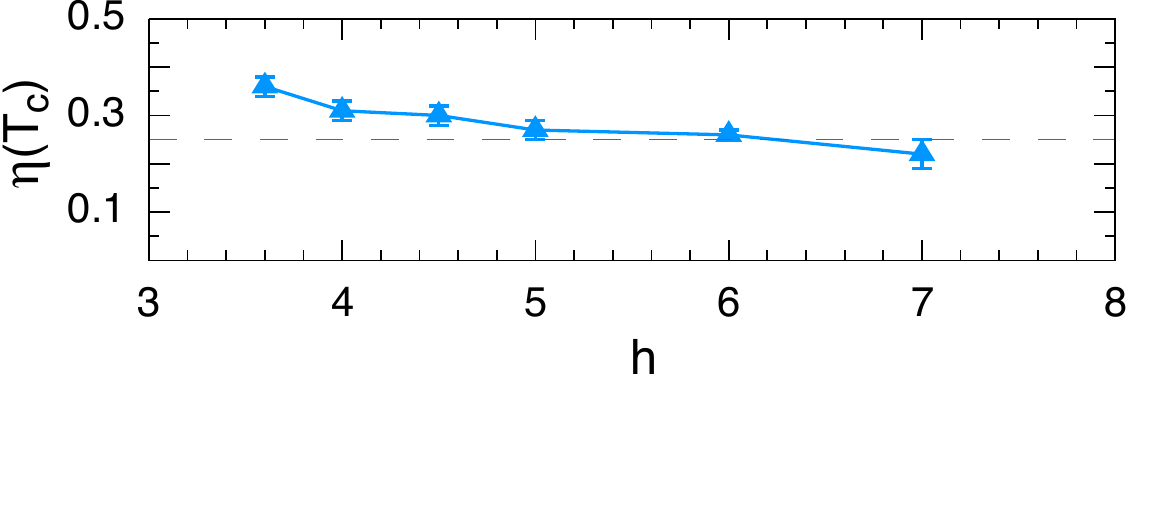}
\end{center}
\caption{\footnotesize{(Color online)
Evolution of the correlation-function exponent $\eta(T_c)$,
related to the $S^x$-$S^y$ spin plane, of the single phase transition between the paramagnet
and 2:1 canted state  as a function of field.
The horizontal dashed line shows the standard BKT value $\eta(T_{\sf BKT})=0.25$.
}}
\label{fig:eta-XY}
\end{figure}

The variation with field of the spin correlations in the $S^x$-$S^y$ plane at this transition are 
also noteworthy, cf. Fig.~\ref{fig:eta-XY}.
The correlation-function exponent $\eta(T_c)$  is observed to also change continuously with
field, albeit more slowly.
We use this value of  $\eta(T_c)$ to find the location of the non-universal jump in spin
stiffness for the other values of magnetic field at $h\gtrsim3.3$ displayed in Fig.~\ref{fig:phase-diagram}.
The critical temperatures thus found are in good agreement with the ones obtained with
the correlation length, taking in account the uncertainties in determining~$\eta$.

Our results strongly suggest the existence of a point at \mbox{$T\approx0.31,h\approx3.2$},
where three apparently \emph{continuous}  transitions meet.
Although simulations approaching this point become very difficult, we find no evidence for any of the transitions becoming first-order.

Lastly, the zero-temperature phase transition from the (collinear) saturated paramagnet into the 2:1
state is observed at a field of $h=9$.
This corresponds to the opening of a gap, at the three-sublattice momenta $\{\mathbf{q}_{\sf K}\}$,
to spin-wave excitations inside the saturated phase.
As the saturation field is approached the required computational effort increases and the accuracy suffers.
However, we also interestingly observe that the agreement between the $T_c$ found by the scaling of
both correlation lengths and  $T_c$ found by the spin-stiffness jump is better at both high and low values
of the  $ 3.3 \lesssim h <9 $ range of applied field.


\section{Discussion and Conclusions}
\label{section:conclusion}



In this paper we have used modern Monte Carlo simulation techniques to explore
the finite-temperature phase diagram of the classical Heisenberg antiferromagnet
on a triangular lattice in applied magnetic field.
The broad outline of this phase diagram has been known for some
decades~\cite{kawamura85}, and all of the phases predicted  --- a collinear one-third
magnetization plateau, together with two algebraically-correlated
coplanar phases, the Y-state and a 2:1 canted phase --- have since been observed
in experiment~\cite{kitazawa99,svistov06,ishii11}.

\begin{table}[t]
\noindent\makebox[-0.2\textwidth]{
\footnotesize
\begin{tabular}{|c |c | c | c | c | c | c | c | c | c |}
\hline
%
%
&
h &
$\mathcal{O}^{zz}(T)$ &
$\chi^{zz} (T)$ &
$\xi^{zz} (T) $ &
${\cal S}^{zz} (r)$ &
${\cal S}^{xy} (r)$ &
{\tiny classification}
\\ \hline
%
%
{\tiny p.m.--plateau} &
{\tiny 2} &
{\tiny $\beta $=$ 1/9$ }&
{\tiny $\gamma $=$ 13/9$ }  &
{\tiny $\nu $=$ 5/6$ }&
{\tiny $\eta $=$ 4/15 $ }&
{\tiny n/a} &
{\tiny 3-state Potts}
\\ \hline
%
%
{\tiny plateau--Y-state} &
{\tiny 2} &
{\tiny n/a} &
{\tiny n/a}  &
{\tiny n/a}  &
{\tiny n/a}  &
{\tiny $\eta(T_{\sf BKT})$=$1/4$} &
\tiny{BKT}
\\ \hline
%
%
{\tiny p.m.--2:1 canted} &
{\tiny 5}   &
{\tiny $\beta $=$ 0.50(5) $} &
{\tiny $\gamma $$\approx$$ 3$}  &
{\tiny $\nu $=$ 2.0(2) $} &
{\tiny $\eta $=$0.50(5)$} &
{\tiny $\eta(T_{c})$=$0.27(2)$} &
{\tiny unknown}
\\ \hline
\end{tabular}
}
\caption{\footnotesize{
Critical exponents for the different continuous phase transitions studied in this paper, as found 
from classical Monte Carlo simulation of the Heisenberg antiferromagnet on a triangular lattice,
at different values of magnetic field $h$.
}}
\label{tab:exponents}
\end{table}

None the less, recent works not withstanding~\cite{gvozdikova11,griset11}, the nature
of the finite-temperature phase transitions between these phases remains surprisingly
poorly understood.
The interest of this problem lies in the fact that both the Y-state and 2:1
canted phase break two qualitatively different symmetries
--- a discrete $\mathds{Z}_3$ symmetry associated with their three-sublattice
structure, and spin-rotation symmetry in $S^x$-$S^y$ plane.
The order parameters for these phases therefore have a compound
$\mathds{Z}_3\otimes O(2)$ character, and it is interesting to ask how
the two symmetries are restored as the temperature is raised. Existing studies of this compound symmetry are very rare\cite{loison04-book,hellmann09}.
The answers we find, summarised in Table~\ref{tab:exponents}, are remarkably
different for the two different phases.


We consider first the case of low values of field $h < 3$.
Here the system exhibits two phases --- a long-range ordered one-third magnetization plateau,
and an algebraically correlated Y-state [cf. Fig.~(\ref{fig:phase-diagram})].
Cooling from the paramagnet reveals two continuous phase transitions,
the expected three-state Potts transition into the one-third plateau, and then
a conventional BKT transition into the Y-state at lower temperature.
Both of these transitions are well-characterised for $h=2$ [Fig.~(\ref{fig:h=2-I})].
However as $h \to 0$ the correlation length in the \mbox{$S^x$-$S^y$} plane
increases dramatically and, for $h\lesssim 0.8$, is comparable with the linear
dimension of the largest clusters simulated for {\it all} temperatures [Fig.~\ref{fig:h=04}(c)].
This effect, combined with the proximity between the three-state Potts and BKT
phase transitions, makes the interpretation of simulation results extremely challenging.
None the less, we are able to obtain a good finite-size scaling of results
for spin-stiffness and $\xi^{zz}$ correlation length down to $h$$\approx $$0.2$,
under the assumption that the two phase transitions remain distinct
and well separated [Fig.~\ref{fig:h=04}(a) and (b)].
It is this, quantitative, analysis of the simulation results which leads to the phase
boundaries shown in Fig.~\ref{fig:phase-diagram}.


At first sight, this result might seem to imply that the double phase transition
--- from paramagnet to plateau, and then from plateau to Y-state --- survives
all the way down to zero field ($h=0$).
We would however council caution : the physics of Heisenberg model in finite
magnetic field may be very different from that in vanishing field, where
the order parameter and excitations have a qualitatively different
character~\cite{kawamura10}.
We note that  exchange anisotropy has recently been argued to act as a
singular
perturbation in the Heisenberg antiferromagnet on the triangular lattice~\cite{misawa10},
and the same role may be played by applied magnetic field for the isotropic
Heisenberg model studied here.
It could also be that proximity to the unconventional phase transition
at $h=0$, renders the finite size scaling used to extract phase boundaries in
Fig.~\ref{fig:phase-diagram} unreliable for $h\to0$.
Previous studies of the Heisenberg antiferromagnet on a triangular lattice
for this range of fields have argued for both a single transition from the
paramagnet into the Y-state~\cite{kawamura85}, and a double transition
of the type described above, but with the temperature window between the two
transitions closing as $h\to0$, see \linecite{gvozdikova11}.
Further simulations with larger cluster sizes, together with a more
sophisticated analysis of results, will be needed to resolve this issue.


Our results for higher fields, $h> 3$, point to a very different scenario.
Here the system undergoes a single, continuous phase transition from the
high-temperature paramagnet into the 2:1 canted phase.
For this to happen, the fluctuations in all three spin components must
become critical at exactly the same temperature.
This would not be unusual in a three dimensional frustrated
magnet~\cite{seabra11}, but it has some very interesting consequences
in the present, two-dimensional model.
Considering first correlations of the $S^z$ components of spin, we find that
the correlation length exponent $\nu$ and order parameter exponent $\beta$
increase with increasing magnetic field [Fig.~\ref{fig:exponents}(a) and (b)].
However, the correlation function exponent remains constant at $\eta\approx 0.5$
[Fig.~\ref{fig:exponents}(c)], a value quite different from three-state Potts transition seen at
low values of field [Fig.~\ref{fig:h=2-II}(c)].
Turning our attention to the $S^x-S^y$ plane, the $\mathcal{O}^{xy}$ order parameter
vanishes in the thermodynamic limit, and spin stiffness shows a jump at the transition
temperature [Fig.~\ref{fig:h5-I}(d)], as would be expected for a BKT transition.
However in this case the correlation-length exponent, $\eta(T_c)$, varies with magnetic field,
and is generically different from the value $\eta(T_{\sf BKT}) =1/4$ found at a BKT
transition [Fig.~\ref{fig:eta-XY}].
This non-universal value of $\eta$ implies a \emph{non-universal} jump in the spin stiffness,
and the temperature at which this jump in the spin stiffness occurs is found to scale as a
polynomial in $1/L$ [inset to Fig.\ref{fig:h5-I}(d)].
These results point to a highly unusual line of continuous phase transitions, interpolating
from a point at which three critical lines meet \mbox{($T\approx0.31,h\approx3.2$)},
to the saturated state at \mbox{($T=0,h=9$)} [cf Fig.~\ref{fig:phase-diagram}].


At an intuitive level, it is easy to see why a phase transition at which $\mathds{Z}_3$ and
$O(2)$ symmetries are broken simultaneously might be different from an isolated
three-state Potts or BKT transition.
The appeal to three-state Potts or BKT universality classes rests on the assumption
of purely short-range interactions.
This condition is unlikely to be met in the combined transition, where critical fluctuations
of one field can mediate a long range interaction for the other.
For example, vortices in the transverse components of spin carry a (topological)
charge and might be expected to couple to $S^z$ components of spins, invalidating
the idea of short-range interactions between these Potts variables.
And, conversely, these vortices can only exist inside a finite-size ``box'' set by the
Potts degree of freedom.
Long range dipolar interactions are well known to induce logarithmic corrections
to scaling in conventional phase transitions~\cite{larkin69}, and it seems
reasonable to suppose that long-range interactions modify the critical exponents
in this compound phase transition.


Precisely what happens where these lines of continuous phase transitions meet, at
\mbox{($T\approx0.31,h\approx3.2$)}, is difficult to say, as it becomes increasing difficult to 
extract reliable estimates of the critical exponents as this point is approached.
Our best estimate of the exponent $\nu$ associated with correlations
of the $S^z$ components of spin tends to $\nu\sim 1$ as $h \to 3.2$ from above
[Fig.~\ref{fig:exponents}(a)].
This is roughly compatible with the three-state Potts value $\nu=5/6$, seen for
the transition from the paramagnet into the one-third magnetization plateau
for $h=2$ [Fig.~\ref{fig:h=2-II}].
However, the order-parameter exponent tends to $\beta\sim 0.2$ [Fig.~\ref{fig:exponents}(b)],
roughly double the three-state Potts value of $\beta=1/9$.
Similarly, considering spin correlations in the $S^x$-$S^y$ plane, we find a
correlation-function exponent $\eta(T_c)\sim0.5$ [Fig.~\ref{fig:eta-XY}],
twice the value observed for the BKT transition from one-third magnetization
plateau to Y-state for $h=2$ [Fig.~\ref{fig:h=2-III}].
Clearly more work needs to be done to understand how the different phases
come together at this point.


Conformal field theory (CFT) has proved to be a very powerful tool for understanding
two-dimensional phase transitions~\cite{itzykson-book}.
Within this approach, every continuous phase transition can be characterised
in terms of a single parameter, the central charge $c$, and critical exponents are
typically rational fractions with discrete values determined by $c$.
In the present case, the three-state Potts transition from the paramagnet to the one-third
magnetization plateau for $h=2$ has central charge $c=4/5$, while the BKT transition
from the one-third magnetization plateau to the 2:1 canted phase for $T=0.05$
has central charge $c=1$.
It has been argued that continuously-varying critical exponents arise most naturally in
Gaussian ($c=1$) CFT's with an additional, marginal, operator~\cite{cardy87}.
However it is hard to see how either a $c=1$ theory, or a direct product of a $c=1$ theory with
another CFT, can be reconciled with the variation of exponents found in our simulations.
We speculate that the transition from the paramagnet to the 2:1 canted state
might therefore provide an example of varying critical exponents associated
with a central charge $c \ne 1$.
This line of phase transitions also provides an example of the concept of  ``weak universality'',
where the exponent $\eta$ is universal while $\beta$ and $\nu$ are allowed to change.
The idea of weak universality was first proposed in the context of the two-dimensional Ising model
with four-spin interaction~\cite{suzuki74}, which can be described by a $c=1$ CFT~\cite{blote86}.
Our results suggest a further generalisation to these ideas to compound phase transitions
which do not necessarily have $c=1$ as the global charge.


It is also instructive to compare these results with existing work on $\mathds{Z}_2\otimes O(2)$
phase transitions in two dimensions.
In principle, systems with $\mathds{Z}_2\otimes O(2)$ symmetry breaking can also
support a line of continuous phase transitions from the disordered state with continuously varying
exponents~\cite{choi84,lee91}.
Recent numerical work suggest that this scenario is not realised in the most
widely studied model, the fully-frustrated XY model on a square lattice~\cite{hasenbusch05}.
However, in a recent twist to the story, Ising and BKT transitions have been observed
to merge into a single phase continuous transition in a more general model~\cite{minnhagen07,minnhagen08}.


The ultimate test of the results contained in this paper would be comparison with the magnetic
phase diagram of a real triangular lattice antiferromagnet.
Here the picture is obscured by terms not present in the isotropic Heisenberg
model, notably magnetic anisotropy and coupling between triangular lattice layers\cite{collins97}.
However, published results for Heisenberg models with easy-axis
anisotropy~\cite{miyashita86}, easy-plane anisotropy~\cite{plumer89}, and interlayer
coupling~\cite{watarai01} suggests that many of the most interesting features
of the phase diagram Fig.~\ref{fig:phase-diagram} survive.
Moreover, the rapid advances in experiments on cold atoms in optical lattices might make
it possible to simulate a truly two-dimensional and isotropic Heisenberg antiferromagnet in the
laboratory~\cite{struck11}.


In conclusion, the behaviour of the Heisenberg model on a triangular lattice in applied
magnetic field is much richer, and much less well understood, than usually supposed.
In this paper we have used modern Monte Carlo simulation techniques to characterise the
different phase transitions which occur as a function of temperature and magnetic field.
The interest of this problems stems from the combined $\mathds{Z}_3 \otimes O(2)$
symmetry of low-temperature coplanar phases.
For values of magnetic field $h \lesssim 3.2 $, we find that the
$\mathds{Z}_3$ symmetry associated with three-sublattice structure
and the $O(2)$ symmetry associated with the spin-rotations in
the $S^x$-$S^y$ plane are broken at different temperatures.
In contrast, for high values of magnetic field $h \gtrsim 3.2 $, we find that
these symmetries are broken at the same temperature,
in a line of continuous phase transitions with continuously varying exponents.
Our results leave a number of important questions unanswered,
including the topology of the phase diagram for $h \to 0$, and the way in which
$\mathds{Z}_3$ and $O(2)$ symmetries combine for
\mbox{($T\approx0.31,h\approx3.2$)}.
Given the importance of finite size effects, it seems unlikely that these questions
can be resolved by simulation alone, without further input from field theory.
We therefore hope that this paper will help to re-open the discussion
of this canonical problem in frustrated magnetism


\section*{Acknowledgments}


The authors thank
Vladimir Dotsentko,
Seiji Miyashita
and Mike Zhitomirsky
for helpful comments on this work, and Matthias Vojta for drawing our attention to some
of the existing literature on the fully-frustrated XY model.
LS acknowledges the hospitality of the Condensed Matter Theory Laboratory
of RIKEN, Wako, where part of this work was completed.
Numerical simulations made use of the Advanced Computing Research Centre,
University of Bristol.
This work was supported by
FCT Grant No.~SFRH/BD/27862/2006,
EPSRC Grants No.~EP/C539974/1 and EP/G031460/1,
and KAKENHI Grants No.~22014016 and No. ~23540397.



\vspace{2cm}

\begin{thebibliography}{99}



\bibitem{wannier50}
G. H. Wannier,
Phys. Rev. \textbf{79}, 357 (1950).

\bibitem{husimi50}
K. Husimi and I. Sy\^ozi,
Prog. Theor. Phys. \textbf{5}, 177 (1950).

\bibitem{anderson73}
P. W.~Anderson,
Mat. Res. Bull. {\bf 8}, 153 (1973).

\bibitem{bernu92} B. Bernu, C. Lhuillier, and L. Pierre,
Phys. Rev. Lett. \textbf{69}, 2590 (1992).

\bibitem{capriotti99}
L. Capriotti, A. E. Trumper and S. Sorella,
Phys. Rev. Lett. \textbf{82}, 3899 (1999).

\bibitem{kawamura84b}
H. Kawamura and S. Miyashita,
J. Phys. Soc. Jpn. \textbf{53}, 4138 (1984).

\bibitem{kawamura93}
H. Kawamura and M. Kikuchi,
Phys. Rev. B \textbf{47}, 1134 (1993).

\bibitem{kawamura98}
H. Kawamura,
J. Phys.: Condens. Mat. \textbf{10}, 4707(1998).

\bibitem{okubo10} T. Okubo and H. Kawamura,
J. Phys. Soc. Jpn. \textbf{79}, 084706 (2010).

\bibitem{kawamura10}
H. Kawamura, A. Yamamoto and T. Okubo,
J. Phys. Soc. Jpn. \textbf{79}, 023701 (2010).

\bibitem{southern93}
B. W. Southern and A. P. Young,
Phys. Rev. B \textbf{48}, 13170 (1993).

\bibitem{southern95}
B. W. Southern and H.J. Xu,
Phys. Rev. B \textbf{52}, R3836 (1995).

\bibitem{wintel95}
M. Wintel, H. U. Everts, and W. Apel,
Phys. Rev. B \textbf{52}, 13480 (1995).

\bibitem{calabrese01} P. Calabrese and P. Parruccini, Phys. Rev. B \textbf{64}, 184408 (2001).

\bibitem{delamotte10}
B. Delamotte, M. Dudka, Y. Holovatch and D. Mouhanna,
Phys. Rev. B \textbf{82}, 104432 (2010).

\bibitem{mekata77}
M.~Mekata,
J. Phys. Soc. Jpn. \textbf{42}, 76 (1977).

\bibitem{kawamura85}
H. Kawamura and S. Miyashita,
J. Phys. Soc. Jpn. \textbf{54}, 4530 (1985).

\bibitem{chubukov89}
A.V. Chubukov and D.I. Golosov,
J. Phys.: Condens. Matter \textbf{3}, 69 (1991).

\bibitem{lee86}
D. H. Lee, J. D. Joannopoulos, J. W. Negele, and D. P. Landau,
Phys. Rev. B \textbf{33}, 450 (1986).

\bibitem{zhitomirsky02}
M. E. Zhitomirsky,
Phys. Rev. Lett. \textbf{88}, 057204 (2002).

\bibitem{gvozdikova11}
M. V. Gvozdikova, P.-E. Melchy and M. E. Zhitomirsky,
J. Phys.: Condens. Mat. \textbf{23}, 164209 (2011).

\bibitem{moliner09}
M. Moliner, D. C. Cabra, A. Honecker, P. Pujol and F. Stauffer,
Phys. Rev. B \textbf{79}, 144401 (2009).


\bibitem{collins97}
M. F. Collins and O. A. Petrenko,
Can. J. Phys. \textbf{75}, 605 (1997).


\bibitem{kitazawa99}
H. Kitazawa, H. Suzuki, H. Abe, J. Tang, and G. Kido,
Physica B: Cond. Mat. \textbf{259-261}, 890 (1999).

\bibitem{svistov06}
L. E. Svistov, A. I. Smirnov, L. A. Prozorova, O. A. Petrenko, A. Micheler, N. B\"uttgen,
A. Y. Shapiro and L. N. Demianets,
Phys. Rev. B \textbf{74}, 024412 (2006).

\bibitem{ishii11}
R. Ishii, S. Tanaka, K. Onuma, Y. Nambu, M. Tokunaga, T. Sakakibara,
N. Kawashima, Y. Maeno, C. Broholm, D. P. Gautreaux, J. Y. Chan and S. Nakatsuji,
Europhys. Lett. \textbf{94}, 17001 (2011).

\bibitem{miyashita86}
S. Miyashita,
J. Phys. Soc. Jpn. \textbf{55}, 3605 (1986).

\bibitem{watarai01}
S. Watarai, S. Miyashita and H. Shiba,
J. Phys. Soc. Jpn. \textbf{70}, 532 (2001).

\bibitem{wang09}
F. Wang, F. Pollmann, and A. Vishwanath,
Phys. Rev. Lett. \textbf{102}, 017203 (2009).

\bibitem{alicea09}
J. Alicea, A. V. Chubukov, and O. A. Starykh,
Phys. Rev. Lett. \textbf{102}, 137201 (2009).

\bibitem{sen09}
A. Sen, F. Wang, K. Damle, and R. Moessner,
Phys. Rev. Lett. \textbf{102}, 227001 (2009).

\bibitem{heidarian10}
D. Heidarian and A. Paramekanti,
 Phys. Rev. Lett. \textbf{104}, 015301 (2010).

\bibitem{seabra10}
L. Seabra and N. Shannon,
Phys. Rev. Lett. \textbf{104}, 237205 (2010).

\bibitem{seabra11}
\mbox{L. Seabra and N. Shannon,
Phys. Rev. B \textbf{83}, 134412 (2011).}

\bibitem{fishman11}
 R. S. Fishman,
 J.  Phys.: Condens. Mat. \textbf{23}, 366002 (2011).

\bibitem{fishman11a}
 R. S. Fishman,
  Phys. Rev. B \textbf{84}, 052405 (2011).

\bibitem{griset11}
C. {Griset}, S. {Head}, J. {Alicea} and O. A. {Starykh},
arXiv:1107.0772 (2011).

\bibitem{stoudenmire09} E. M. Stoudenmire, S. Trebst and L. Balents, Phys. Rev. B \textbf{79}, 214436 (2009).

\bibitem{melchy09}
P.-E. Melchy and M. E. Zhitomirsky,
Phys. Rev. B \textbf{80}, 064411 (2009).

\bibitem{mermin66} 
N. D. Mermin and H. Wagner, 
Phys. Rev. Lett. \textbf{17}, 1133 (1966).

\bibitem{berezinskii72}
V. L. Berezinskii,
Sov. Phys. JETP \textbf{32}, 493 (1971).

\bibitem{kosterlitz73}
J. M. Kosterlitz and D. J. Thouless,
J. Phys. C Solid State \textbf{6}, 1181 (1973).

\bibitem{villain77}
J. Villain,
J. Phys. C {\bf 10}, 4793 (1977).

\bibitem{miyashita84}
S. Miyashita and H. Shiba,
J. Phys. Soc. Jpn. {\bf 53}, 1145 (1984).

\bibitem{teitel83-PRL51}
S. Teitel and C. Jayaprakash,
Phys. Rev. Lett. {\bf 51}, 1999 (1983).

\bibitem{choi84}
M. Y. Choi and S. Doniach,
Phys. Rev. B \textbf{31}, 4516 (1985).

\bibitem{lee91}
J. Lee, E. Granato, and J. M. Kosterlitz,
Phys. Rev. B \textbf{44}, 4819 (1991).

\bibitem{loison04-book}
D. Loison,
in \emph{Frustrated Spin Systems},
edited by H. T. Diep (World Scientific, Singapore, 2005).

\bibitem{hasenbusch05}
M. Hasenbusch, A. Pelissetto and E. Vicari,
J. Stat. Mech., P12002 (2005).

\bibitem{cristofano06}
G. Cristofano, V. Marotta, P. Minnhagen, A. Naddeo and G. Niccoli,
J. Stat. Mech., P11009 (2006).

\bibitem{minnhagen07}
P. Minnhagen, B. J. Kim, S. Bernhardsson, and G. Cristofano,
Phys. Rev. B \textbf{76}, 224403 (2007).

\bibitem{minnhagen08}
P. Minnhagen, B. J. Kim, S. Bernhardsson, and G. Cristofano,
Phys. Rev. B \textbf{78}, 184432 (2008).

\bibitem{kawamura84-JPSJ53}
H. Kawamura,
J. Phys. Soc. Jpn. \textbf{53}, 2452 (1984).

\bibitem{matsuda70}
H. Matsuda and T. Tsuneto,
Prog. Theor. Phys. Supp. \textbf{46}, 411 (1970).

\bibitem{liu73}
K. S. Liu and M. E. Fisher,
J. Low Temp. Phys. {\bf 10}, 655 (1973).

\bibitem{tay10}
T. Tay and O. I. Motrunich,
Phys. Rev. B \textbf{81}, 165116 (2010).

\bibitem{hukushima96}
K. Hukushima and K. Nemoto,
J. Phys. Soc. Jpn. \textbf{65}, 1604 (1996).

\bibitem{nelson77}
D. R. Nelson and J. M. Kosterlitz,
 Phys. Rev. Lett. \textbf{39}, 1201 (1977).


\bibitem{weber88}
H. Weber and P. Minnhagen,
Phys. Rev. B \textbf{37}, 5986 (1988).

\bibitem{binder81}
K. Binder, 
Z. Phys. B \textbf{43}, 119 (1981).


\bibitem{wu82}
F.~Y.~Wu,
Rev. Mod. Phys. {\bf 54}, 235 (1982).

\bibitem{hellmann09} 
M. Hellmann, Y. Deng, M. Weiss, and D. W. Heermann,
J. Phys. A-Math. Theor. \textbf{42}, 225001 (2009).

\bibitem{misawa10}
T. Misawa and Y. Motome,
J. Phys. Soc. Jpn. \textbf{79}, 073001 (2010).


\bibitem{larkin69}
A. I. Larkin and D. E. Khamel'nitskii,
Sov. Phys. JETP \textbf{29}, 1123 (1969).

\bibitem{itzykson-book}
C. Itzykson and J.M. Drouffe,
\emph{Statistical Field Theory} (Cambridge University Press,~1989)~Vol.~2.

\bibitem{cardy87}
J. L. Cardy,
J.  Phys. A: Math.  Gen. \textbf{20}, L891 (1987).

\bibitem{suzuki74}
M. Suzuki,
Prog. Theor. Phys. \textbf{51}, 1992 (1974).

\bibitem{blote86}
H. W. J. Bl\"ote, J. L. Cardy and M. P. Nightingale,
Phys. Rev. Lett. \textbf{56}, 742 (1986).

\bibitem{plumer89}
M. L. Plumer, A. Caill\'e and K. Hood,
Phys. Rev. B \textbf{39}, 4489 (1989).

\bibitem{struck11}
 J. Struck, C. Ölschläger, R. Le Targat, P. Soltan-Panahi, A. Eckardt, M. Lewenstein, P. Windpassinger, and K. Sengstock, 
 Science \textbf{333}, 996 (2011).





\end{thebibliography}
\end{document}